\documentclass{aa}

\usepackage[varg]{txfonts}
\usepackage{siunitx}
\usepackage[version=4]{mhchem}
\usepackage{multirow}
\usepackage{graphicx}
\usepackage[caption=false]{subfig}
\usepackage[strict]{changepage}
\usepackage{xcolor}
\usepackage{ulem}
\usepackage[colorlinks=true,citecolor=blue]{hyperref}
\usepackage{float}
\usepackage{amssymb}
\usepackage{amsmath}
\usepackage{lscape}
\usepackage{mathtools}
\usepackage{natbib}
\usepackage{pdflscape}
\usepackage{gensymb}
\usepackage[tableposition=top]{caption}

\DeclareSIUnit\jansky{Jy}

\title {High Resolution LAsMA $^{12}$CO and $^{13}$CO Observation of the G305 Giant Molecular Cloud Complex : I. Feedback on the Molecular Gas.}

\author{P. Mazumdar\inst{\ref{inst1}}
\and F. Wyrowski \inst{\ref{inst1}} 
\and D. Colombo \inst{\ref{inst1}}
\and J.\,S.\,Urquhart \inst{\ref{inst2}}
\and M.\,A.\,Thompson \inst{\ref{inst3}}
\and K. M. Menten\inst{\ref{inst1}}
}

\institute{Max-Planck-Institut f\"{u}r Radioastronomie, Auf dem H\"{u}gel 69, 53121 Bonn, Germany \label{inst1} \\
\email{pmazumdar@mpifr-bonn.mpg.de}
\and
Centre for Astrophysics and Planetary Science, University of Kent, Canterbury, CT2 7NH, UK \label{inst2}
\and
Centre for Astrophysics Research, Science and Technology Research Institute, University of Hertfordshire, College Lane, Hatfield AL10 9AB, UK \label{inst3}
}

\date{Received 22.12.2020 / Accepted 21.05.2021}

\abstract{Understanding the effect of feedback, the interaction of young  massive stars with their parental Giant Molecular Clouds, is of central importance for studies of the interstellar medium and star formation.}{We observed the G305 star forming complex in the $J=3\text{-}2$ lines of \ce{^{12}CO} and \ce{^{13}CO} to investigate how molecular gas surrounding the central stellar clusters is being impacted by feedback.}{The APEX telescope's LAsMA multi-beam receiver was used to observe the region. Excitation temperatures and column density maps were produced. Combining our data with data from the SEDIGISM survey resulted in a \ce{^{13}CO} $J=3\text{-}2/2\text{-}1$ excitation map. To verify whether feedback from stellar clusters is responsible for exciting the gas, the distribution of CO excitation was compared with that of 8$\,\mu\rm{m}$ emission imaged with Spitzer, which is dominated by UV-excited emission from PAHs. Line centroid velocities, as well as stacked line profiles were examined to investigate the effect of feedback on the gas dynamics.}{Line profiles along radially outward directions demonstrate that the excitation temperature and \ce{^{13}CO} $J=3\text{-}2/2\text{-}1$ ratio increase steeply by factors of $\sim$\,2--3 at the edge of the denser gas traced by \ce{^{13}CO} that faces the hot stars at the center of the complex and steadily decreases away from it. Column density also increases at the leading edge, but does not always decrease steadily outward. Regions with higher 8$\,\mu\rm{m}$ flux have higher median excitation temperatures, column densities and \ce{^{13}CO} $J=3\text{-}2/2\text{-}1$ ratio. The centroid velocity probability distribution function of the region shows exponential wings, indicative of turbulence driven by strong stellar winds. Stacked spectra in regions with stronger feedback have  higher skewness and narrower peaks with pronounced wings compared to regions with weaker feedback.}{Feedback from the stellar cluster in G305 has demonstrable effects on the excitation as well as on the dynamics of the giant molecular cloud.}


\begin{document}

\titlerunning{G305 Giant Molecular Cloud : I. Feedback on Molecular Gas}
\authorrunning{P. Mazumdar et al.}
\maketitle

\section{Introduction} \label{sec:intro}
    Massive stars ($M>8\si{M_{\odot}}$) are rare and usually form inside giant molecular clouds (GMCs) as the dominant members of young stellar clusters \citep{Motte2018}. They are short lived ($\leq 30\,\si{Myr}$), but are known to inject large amounts of feedback into the interstellar medium in the form of stellar winds, ionizing radiation and supernovae \citep{krumholz2014}. These feedback mechanisms can in turn trigger or disrupt the formation of the next generation of stars when they interact with the natal molecular cloud. They can sweep up the surrounding gas and create parsec-scale cavities around them \citep{deh}, forming dense shells of gas as a result and thereby triggering star formation. Conversely, they can also completely disperse their surrounding molecular gas suppressing star formation \citep[and references therein]{krumholz2014}. The ability to both constructively and destructively affect the formation of subsequent generation of stars means that high mass stars play a significant role in driving the evolution of GMCs \citep{zin07}. Here, we study the feedback of massive stars in the G305 HII region and molecular cloud complex.
     
    \begin{figure*}[t]
        \centering
        \includegraphics[angle=0, origin=c, trim = 0cm 11.5cm 0pt 0pt,clip,width=18cm]{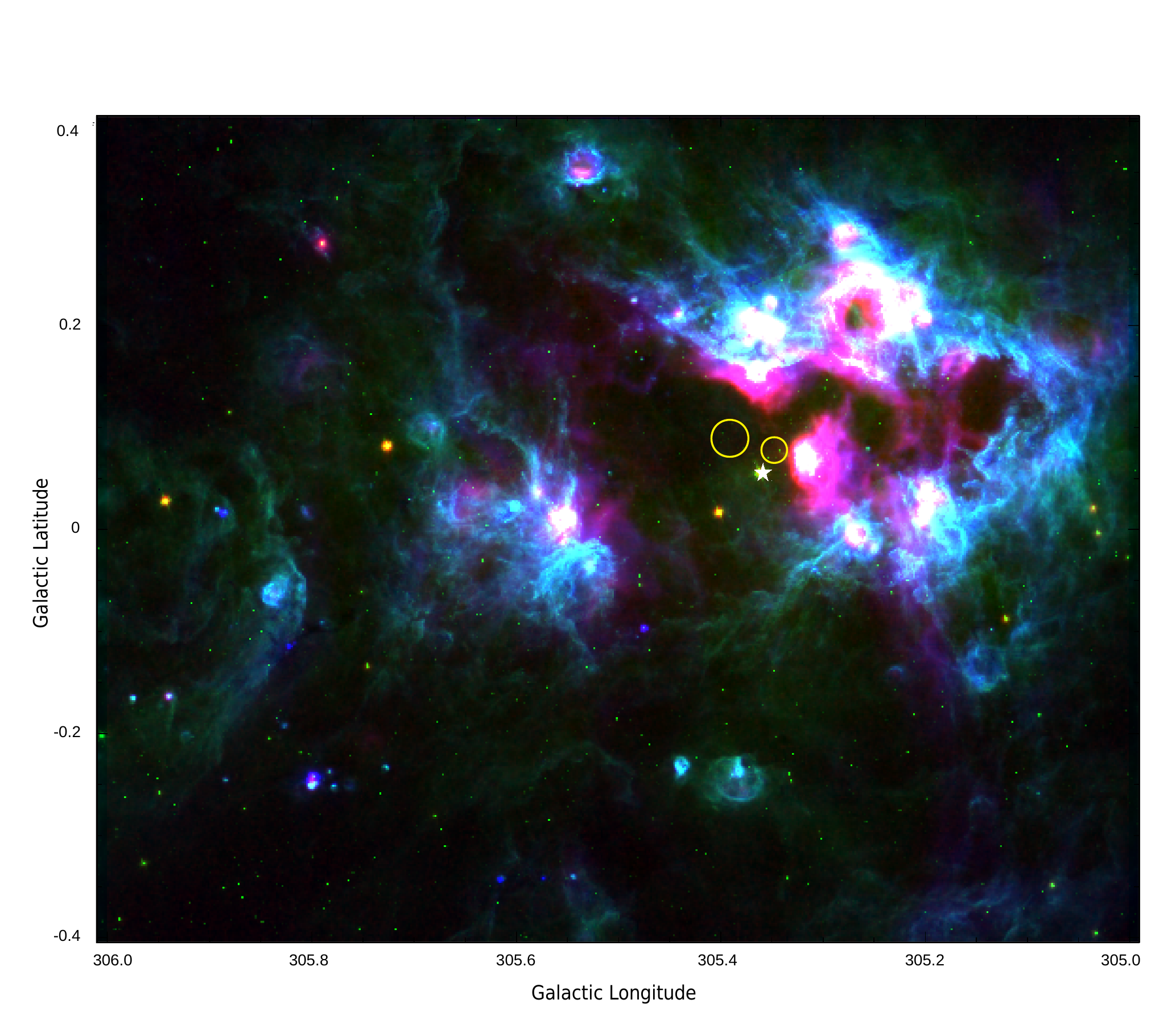}
        \caption{Three-color image (green = \textit{Spitzer}-IRAC4 8\,$\mu \rm{m}$, red = Midcourse Space Experiment (MSX) 21.3\,$\mu \rm{m}$, blue = \textit{Herschel}-PACS 70\,$\mu \rm{m}$) of the G305 star-forming complex. The 21.3\,$\mu \rm{m}$ emission is dominated by hot dust in the HII region. The colder gas is traced by the 70\,$\mu \rm{m}$ emission. The interface between the ionized and molecular gas appears as a blend of green (strong 8\,$\mu \rm{m}$ emission from PAHs), blue (colder molecular gas) and occasionally red (interfaces very close to HII regions). The positions of Danks 1 and 2 clusters have been marked with the smaller and the larger yellow circles respectively and the wolf-rayet star WR48a has been marked as a filled white star.}
        \label{fig:intro-fig}
    \end{figure*}
    
    \subsection*{G305: a brief history} \label{sec:history}
        The Giant Molecular Cloud associated with G305 is one of the most massive and luminous in the Milky Way (Fig. \ref{fig:intro-fig}). It is located in the Galactic plane at $l\sim\ang{305}\,,\, b\sim\ang{0}$ and at a kinematic distance of $4\,\si{kpc}$ (derived from a combination of radio and H$\alpha$ observations by \citet{clark}; \citet{davies12} measured its spectrophotometric distance to be $3.8\pm 0.6\,\si{kpc}$ and most recently \citet{borissova} measured the \textit{Gaia} DR2 average distance to be $3.7\pm1.2\,\si{kpc}$), which places it in the Scutum-Crux spiral arm. Given this distance, the complex has a diameter of $\sim 30\,\si{pc}$ \citep{clark} and a molecular mass of $\sim 6\times10^5\,\rm{M_{\odot}}$ \citep{hindson10}. The G305 complex consists of a large central cavity that has been cleared by the winds from massive stars belonging to two visible central clusters (Danks 1 and 2) and the Wolf-Rayet star (WR48a) \citep{clark, davies12} and is surrounded by a thick layer of molecular gas \citep[traced by CO and NH$_3$ emission;][]{hindson10,hindson13}. Radio continuum observations by \citet{hindson12} have revealed that the cavity is filled with ionised gas and identified 6 ultra-compact HII (UC HII) regions and also one bright rimmed cloud (BRC) at the periphery of the cavity, indicating molecular gas irradiated by UV radiation \citep{Sugitani1994, thompson2004}, that may cause implosion \citep{bertoldi1989} or evaporation. A number of studies have reported star formation tracers (water and methanol masers, HII regions and massive young stellar objects, MYSOs) \citep{clark,lumsden2013,urquhart2014rms,green09,green_MMB}. Furthermore, \citet{hindson10} found the concentration of star formation tracers to be enhanced inside a clump of NH$_3$ bearing molecular gas that faces the ionising sources, which is consistent with the hypothesis that the star formation has been triggered. Analysis of the stellar clusters in the complex reveals them to have ages of $1.5\,\si{\mega yr}$ for Danks 1 and $3\,\si{\mega yr}$ for Danks 2, with the former possibly being triggered by the latter \citep{davies12}. Additionally, a diffuse population of evolved massive stars was also found to exist within the confines of the G305 complex that had formed around the same time as the two clusters \citep{leistra2005,shara, mauerhan,davies12,faimali,borissova}.

        \par How feedback from the massive stars affects molecular clouds is still poorly understood. However, the extensive amount of work done in identifying and characterising the ongoing star formation taking place in this complex (UC HIIs, deeply embedded protostars and protoclusters) and mapping the distribution of molecular and ionized gas, makes G305 the ideal laboratory to study the role of feedback in affecting molecular cloud structures and creating future generation of stars. We have divided our work into two separate parts. In this paper, we want to further our understanding of how feedback from the central population of stars and the UC H {\small {II}} regions in G305 have affected the molecular gas. In the second paper of this series we will decompose the GMC into clumps and investigate whether any discernible differences exist in properties of clumps experiencing feedback compared to those further away from the feedback region.
        
        \par So far, the evidence of feedback on the molecular gas in G305 has mostly been phenomenological and qualitative. In this paper, we study the excitation and kinematics of the gas in the complex since that is the most direct evidence of the feedback affecting the molecular cloud. \citet{hindson13} observed the $J=1\text{-}0$ line of the \ce{CO} main isotopologue to trace the distribution of molecular gas with a resolution of $> 30 \si{\arcsecond}$. Since, the compressed layer of gas due to the feedback is expected to be thin, resolving it requires a higher angular resolution. The feedback region will also exhibit higher excitation of gas that needs to be verified with excitation studies. In addition, the \ce{^{12}CO} becomes optically thick for moderately dense clouds ($\sim 10^3\,\si{\per\cubic\cm}$ at $25\,\si{\kelvin}$ for the ($J=1\text{--}0$) transition) and consequently, provides unreliable measurements. This necessitates the need for observations with rarer isotopologues of \ce{CO}. Finally, observations of the dispersed gas requires sensitivity to very low column densities. Therefore, we decided to use the Large APEX sub-Millimeter Array (LAsMA) seven pixel array on the Atacama Pathfinder EXperiment (APEX) 12 meter sub-millimeter telescope to observe the G305 complex in the $J=3\text{-}2$ rotational transition of both \ce{^{12}CO} and its \ce{^{13}CO} isotopologue.
        
        \par The paper is organized as follows. We describe the observation methods and data reduction techniques in Sect. \ref{sec:obs-red}. We then present the results and the following analyses of the data along with their discussions in Sects. \ref{sec:res} - \ref{sec:dynamics}. In Sect. \ref{sec:conc} we summarize the findings of this paper.

\section{Observation and Data Reduction \label{sec:obs-red}}
    
    \subsection{Observations}
    We mapped a $1\degree \times 1\degree$ area centred on $(l,b)=(305.0,0.0)$. The observations were done between 2017 and 2019 using the APEX telescope \citep{APEX} under the project number M-099.F-9527A-2017. The APEX telescope is located at an altitude of $5100 \, \si{\m}$ at Llano de Chajnantor, in Chile.\footnote{\tiny This publication is based on data acquired with the Atacama Pathfinder EXperiment (APEX). APEX is a collaboration between the Max-Planck-Institut f\"{u}r Radioastronomie, the European Southern Observatory and the Onsala Space Observatory.}

    \par The receiver employed for these observations was the 7 pixel LAsMA MPIfR PI instrument operating in the frequency range of $270-\SI{370}{\GHz}$ \citep{APEX-arrays}. It is a hexagonal array of six pixels surrounding a central pixel. The outer array is separated from the central pixel by $\sim 2$ FWHM. The pixels employ superconductor-insulator-superconductor (SIS) sideband separating mixers (2SB). The backend consisted of Fast Fourier Transform Spectrometers (FFTS4G) that cover an intermediate frequency (IF) bandwidth of 4--8 $\si{\GHz}$ instantaneously with 65536 spectral channels with a width of $\SI{61.03}{\kHz}$.

    \par In order to observe both $\ce{^{13}CO}$ ($\nu_{\text{rest}}\sim 330.588\,\si{\GHz}$) and $\ce{^{12}CO\,(3-2)}$ ($\nu_{\text{rest}}\sim345.796\,\si{\GHz}$) simultaneously the local oscillator frequency was set at $\SI{338.190}{\GHz}$. This frequency was chosen in order to avoid contamination of the $\ce{^{13}CO}\,(3\text{-}2)$ lines due to bright $\ce{^{12}CO}\,(3\text{-}2)$ emission from the image band. For this setup, the FWHM is $\sim \SI{19}{\arcsecond}$, and the velocity resolution is $\SI{0.053}{\km\per\second}$. The whole 1 square degree region was divided into four sub-maps -- one for each quadrant centered around $l=\ang{305.5;;} \, , \,b=\ang{0;;}$. Each sub-map was further divided into 9 smaller sub-maps of size $\SI{10}{\arcmin} \times \SI{10}{\arcmin}$ each. Observations were performed in a position switching on-the-fly (OTF) mode. The reference-positions were carefully chosen and tested to ensure no emission was present in the velocity range of $-150\, \text{to}\, \SI{50}{\km\per\s}$ (see Sect. \ref{app:a} for details about the reference positions). For each scan, the 7 pixel-array was rotated to an angle of $\ang{19.1}$ in equatorial coordinates to ensure minimal overlap between the pixels and maximize sky coverage. For each OTF scan, data were dumped every $\SI{7}{\arcsecond}$ scanned. On reaching the edge of every mapped area, the array was shifted perpendicular to the scan direction by half the beam-width ($\SI{9}{\arcsecond}$) and the scan was then carried out in the reverse direction. This process was repeated until the whole sub-map was observed. The same sub-map was then observed in a $\ang{90}$ rotated frame to ensure Nyquist sampling was achieved along both directions and also to reduce systematic scanning effects. During each scan, a calibration measurement was made every 10-15 minutes. This setup of a sample spacing of $\sim \SI{7}{\arcsecond}$ along with a sampling time of $\SI{0.1}{\sec}$ resulted in datacubes of size $10\si{\arcmin} \, \times \, 10\si{\arcmin}$ in approximately 1 hour.
    \par The calibration of the spectral lines was carried out using the \texttt{apexOfflineCalibrator} pipeline \citep{dirk2006,apex_calibration}. It uses a three load chopper wheel method, which is an extension of the ``standard'' method used for millimetre observations \citep{ulich} to calibrate the antenna temperature scale. To remove the spectral variations of the atmosphere across the bandpass, the ATM model \citep{pardo} was used. The pipeline provides the flexibility to determine the opacity for the whole band pass as one number or to calculate a more accurate opacity channel by channel. We chose the latter for calibrating our data since the \ce{^{13}CO} line lies at the edge of an atmospheric water line at 325~GHz and hence an accurate accounting for this line in the bandpass is essential. After the calibration, the intensities are obtained on the $T^*_A$ (corrected antenna temperature) scale. Apart from the atmospheric attenuation, this also corrects for rear spillover, blockage, scattering and ohmic losses. All intensities stated in this paper are in the $T^*_A$ scale, unless specifically stated otherwise. Regular observations of Jupiter were done during the commissioning of the instrument and the observations of G305, resulting in a beam efficiency of $\eta_{mb}=0.74$ with an uncertainty of about $10\%$. This efficiency value was used to convert intensities from $T^*_A$ scale to the main beam brightness temperature ($T_{mb}$).

    \subsection{Data Reduction}
    
    The data was reduced using the GILDAS package\footnote{\url{http://www.iram.fr/IRAMFR/GILDAS}}. A velocity range of $\SI{-200}{\km\per\s}$ to $\SI{200}{\km\per\s}$ was extracted from every spectrum and resampled to an adequate velocity resolution of $\SI{0.5}{\km\per\s}$ to reduce the noise. The velocity range $\SI{-150}{\km\per\s}$ to $\SI{50}{\km\per\s}$ was masked before fitting a 3rd order baseline to each spectrum. The reduced, calibrated data obtained from the different scans were then combined and gridded using a $6\si{\arcsecond}$ cell size. The gridding process includes a convolution with a Gaussian kernel with a FWHM size of one-third the telescope FWHM beam width. The data cubes obtained have a final angular resolution of $\sim 20\si{\arcsecond}$. Spectra from all the pixels were then averaged to find the portions of the spectra containing line emission, which were then masked. The process of baseline removal was again repeated for the spectra of each pixel to obtain a more stable baseline and a cleaner map.

        \begin{landscape}
            \begin{figure}
                \includegraphics[width=0.65\textwidth]{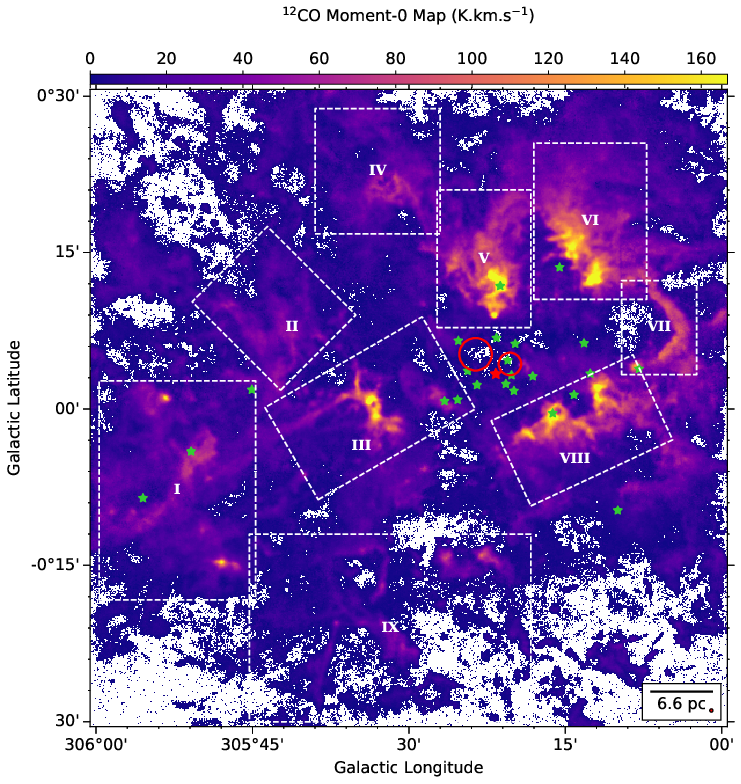}
                \includegraphics[width=0.65\textwidth]{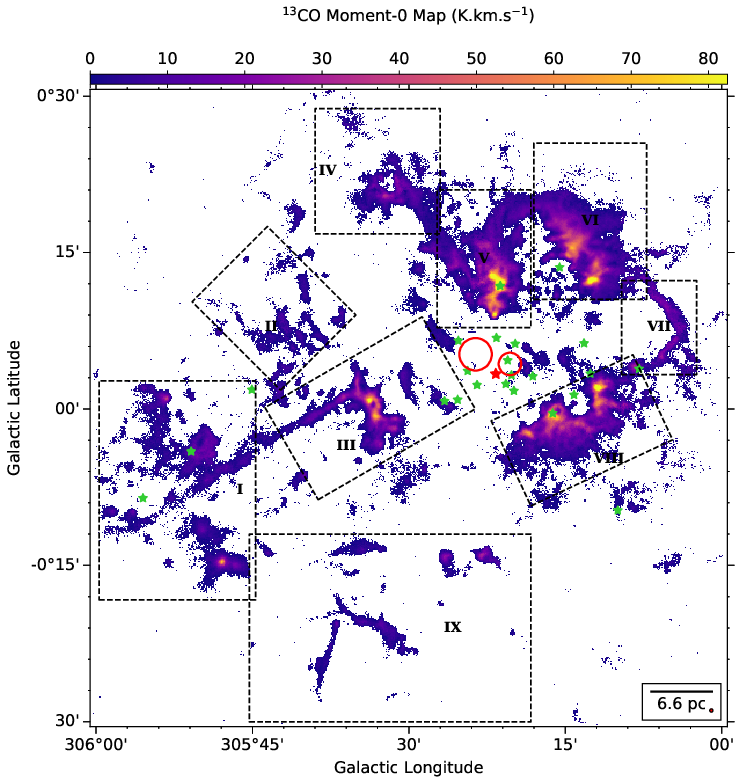}
                \caption{Moment-0 maps of \ce{^{12}}CO (3\text{-}2) [\textit{left}] and \ce{$^{13}$}CO (3\text{-}2) [\textit{right}] lines towards the G305 GMC complex. In both maps, the emission has been integrated over a velocity range from $-70$ to $+10\, \si{\km\per\s}$. Pixels with no emission $>5\sigma$ for at least three consecutive velocity channels have been excluded (seen in white). Overlaid on top as green stars are the position of stars as reported in \citet{borissova}. The red circles show the positions of the Danks 1 and 2 clusters and the red star is WR48a. The numbered regions are discussed in the text.}
                \label{fig:mom-0}
            \end{figure}
        \end{landscape} 

\section{Integrated Properties \label{sec:res}}

    \subsection{Moment-0 Maps} \label{sec{int-prop}}
        In order to create the integrated intensity (``moment-0'') maps of the G305 region, a noise map of each grid-element (referred to as pixel from now onwards) was first created for both the \ce{^{12}CO} and \ce{^{13}CO} $J=$ 3\text{-}2 transitions. This was done by calculating the standard deviation for a part of the spectrum consisting of 100 emission free channels. Fig. \ref{fig:noise_map} shows the pixel-wise noise of the whole region in both \ce{^{12}CO} and \ce{^{13}CO} $J=$ 3\text{-}2 data sets. The moment-0 map was produced by integrating the emission above 5$\sigma$ between $\SI{-70}{\km\per\second}$ and $\SI{+10}{\km\per\second}$. Fig. \ref{fig:mom-0} shows our moment-0 maps of the G305 star forming complex. The \ce{^{12}CO} map traces the large scale, diffuse as well as hot gas very well.

         The \ce{^{13}CO} map on the other hand is sensitive to higher column density gas and hence, traces the more compact, dense clumps in the complex\footnote{Due to the high opacity of \ce{^{12}CO}  and a resulting much lower effective critical density, the \ce{^{13}CO}  transition probes significantly higher \ce{H2} volume densities in the clouds}. We have divided the map into nine regions for ease of explanation. This demarcation of the regions was done visually to highlight relevant molecular gas features observed in the channel-maps. From the two integrated maps it can be seen that the central region of the giant molecular cloud has been cleared out of most of the high density gas. Although it is reasonable to believe that the central stars clusters are responsible for this, it still needs to be verified whether the amount of feedback from the stars in the complex can indeed carve out such a large hole. This will be explored in Sect.\ref{sec:feedback_effects}.
        
        Looking at the gas content of the complex, we observe three bright dense molecular cloud complexes. One in the (Galactic) north (region V and VI in Fig.\ref{fig:mom-0}), one in the south (region VIII) and the third in the east (region III) of the central cavity. The northern and the southern complex are connected by a thin strip of high(er) column density tenuous gas (region VII) towards the west. Both of these complexes also show a carved-out hole in their center where one O and two B stars have been observed \citep{leistra2005,borissova}. The cloud to the east (III) is disconnected from the others but has a long very straight ($\sim20\,\times\,2\,$pc) filamentary structure (III and I) trailing away from the cavity itself. It has a central velocity of $\sim 40\,\si{\km\per\s}$ and a gradient of $\sim 0.25\,\si{\km\per\second}\text{pc}^{-1}$. Such filaments are found to be associated with many star-forming regions and are believed to play a crucial role in star-formation \citep{andre}. The north-east of the cavity appears as a wind-blown structure delineated by regions II, III, IV and V. Towards the south of the cavity (between regions III and VIII), the gas also appears to have been dispersed by the feedback.

    \subsection{Velocity structure} \label{sec:velo-str}

\begin{figure}[h]
            \centering
            \hspace{-1cm}
            \includegraphics[width=0.35\textwidth]{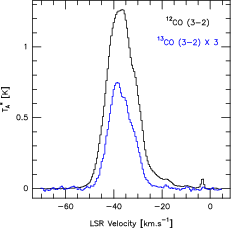}
            \caption{Averaged spectrum of \ce{^{12}CO} and \ce{^{13}CO} over the whole G305 giant molecular cloud complex. The \ce{^{13}CO} line has been scaled up by a factor of 3.}
            \label{fig:spectra}
        \end{figure}

        Fig.\,\ref{fig:spectra} shows the average spectrum over the whole region.  There is no confusion in the foreground except for some local emission at $\sim -4\,\si{\km\per\s}$ in \ce{^{12}CO}. The line profile is not Gaussian. All emission from the complex is found within the velocity range between $-50$ and $-26\, \si{\km\per\s}$ for \ce{^{13}CO} and $-55$ and $-7\, \si{\km\per\s}$ for \ce{^{12}CO}.

        \begin{figure*}[ht]
            \centering
            \includegraphics[trim = 3cm 2cm 11.7cm 1cm, clip, width=1.0\textwidth]{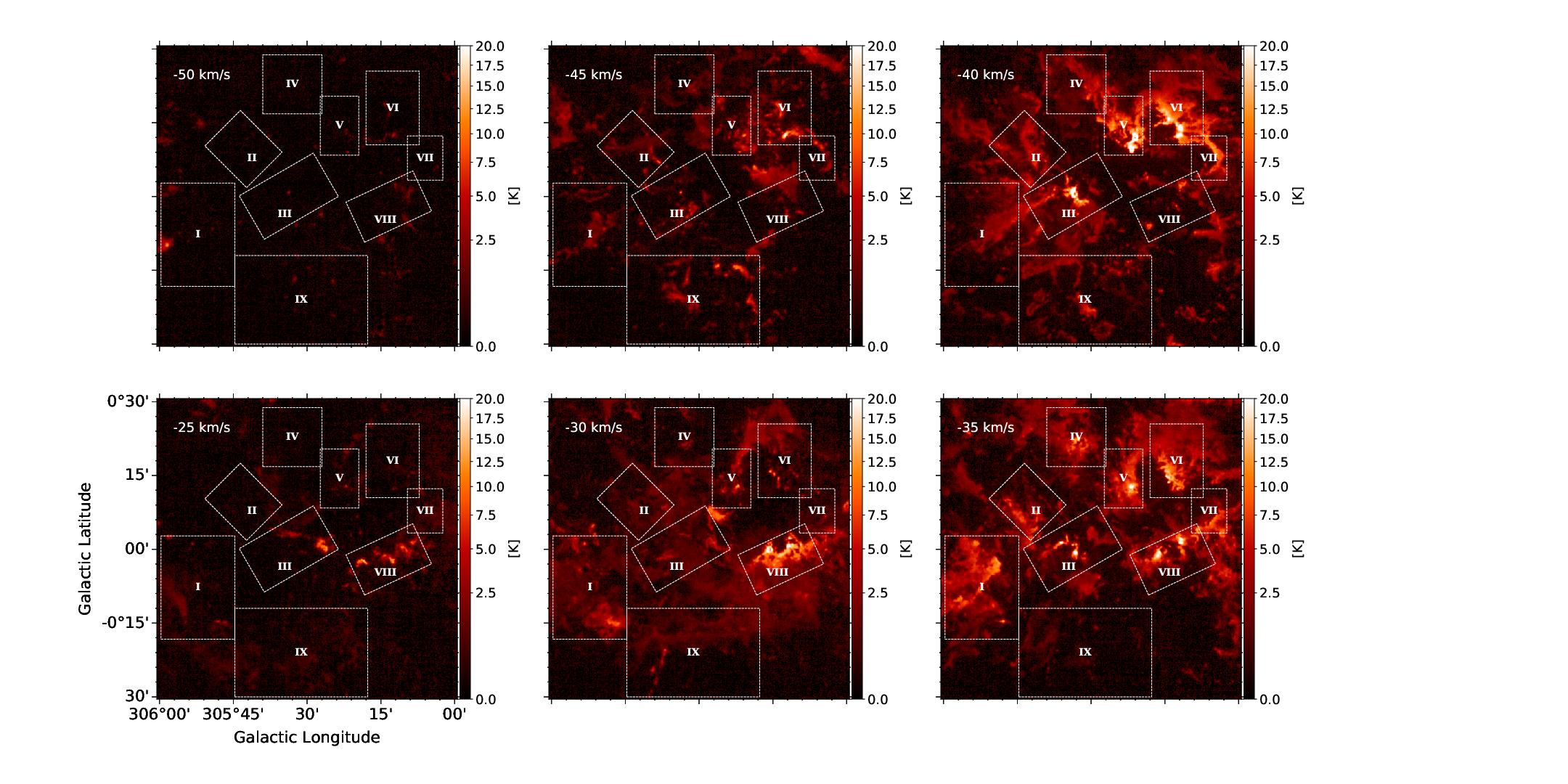}
            \caption{Channel maps of \ce{^{12}CO}(3\text{-}2) emission towards G305. The emission has been integrated over $5\,\si{\km\per\s}$ around the velocities indicated in each panel.}
            \label{fig:chan-map}
        \end{figure*}

        \par Fig. \ref{fig:chan-map} shows the channel maps of G305 for the range of velocities $-50 \rm{\ to} -25\,\si{\km\per\s}$ with a spacing of $5\,\si{\km\per\s}$. Regions I and VIII are prominent in the range $-25$ to $-35\,\si{\km\per\s}$, while the other regions (II, III, IV, V, VI and VII) emit in the range $-35$ to $-45\,\si{\km\per\s}$. The observed velocity structure is most likely a consequence of the interplay of the morphology of the cloud and the feedback from the central star clusters on the natal molecular cloud. For a comprehensive analysis of the morphology of the cloud we refer to \citet{hindson13} who studied the region reported here in more detail by comparing the molecular emission with the morphology of the surfaces of molecular clouds illuminated by the far-ultraviolet (FUV) photons of the stellar sources, commonly refered to as photon dominated regions (PDR) \citep{Tielens1985}. Our results are consistent with the picture presented by these authors.
        
\section{LTE Analysis \label{sec:lte}}
        The CO data were combined to calculate the excitation temperature, optical depth, and column density. We assume that the molecular gas can be described as a system in Local Thermodymanic Equilibrium (LTE) and the brightness temperature ($T_B$) is equal to the measured $T_{mb}$. Then, solving the radiative transfer equation for an isothermal slab of CO radiating at a frequency $\nu$ we obtain:
        \begin{equation}\label{eq:LTE-1}
            \frac{T_B}{\si{\kelvin}} = [J_{\nu}(T_{ex})-J_{\nu}(T_{bg})] \cdot \left(1-e^{-\tau_{\nu}}\right)
        \end{equation}
        \newline where $T_{ex}$ is the excitation temperature of the line; $T_{bg}$ is the temperature of the cosmic microwave background; $\tau_{\nu}$ is the optical depth; and, $J_{\nu}(T)$ is the equivalent temperature of a black body at temperature $T$ which can be written as:
        \begin{equation}\label{eq:LTE-2}
            \frac{J_{\nu}(T)}{\si{\kelvin}} = \frac{h\nu}{k_B} \left( \frac{1}{e^{h\nu/k_BT}-1}\right)
        \end{equation}
        \newline $h$ is the Planck's constant and $k_B$ is the Boltzmann constant.

        \par All the calculations were done on each $l,b,v$ three-dimensional pixel (from here on referred to as voxel). The advantage of this methodology over that of deriving properties over the velocity integrated values is that all the subsequent properties derived are independent of any segmentation method used for source extraction.

        \par To calculate the $^{12}$CO excitation temperature, we adopt the textbook formalism \citep{wilson_book}, which assumes that the $^{12}$CO (3\text{-}2) line is optically thick. This was then used to determine the $^{13}$CO optical depth and subsequently the $^{13}$CO column density. Eq. \ref{eq:LTE-1} is used to calculate the excitation temperature. Using a value of $2.7\,\si{\kelvin}$ for the (cosmic microwave radiation) background temperature, we obtain:
        \begin{equation}\label{eq:LTE-3}
          \frac{T_{ex}}{\si{\kelvin}} = 16.6 \, \left[\ln{\left(1+\frac{16.6}{T_{12}+0.04}\right)}\right]^{-1},
        \end{equation}
        where we use $T_{12}$ and $T_{13}$ for the main-beam brightness temperature of $^{12}$CO and $^{13}$CO, respectively. All the voxels satisfied the condition $T_{12} > T_{13}$; so, we did not have any significant case of self-absorption in our data that would have rendered the excitation temperature derived from this method unreliable.
        
        \par The optical depth is derived for each voxel from the excitation temperature obtained from $^{12}$CO and the $^{13}$CO intensity using the following equation derived from Eq. \ref{eq:LTE-1}:
        \begin{equation} \label{eq:LTE-4}
            \tau_{13} = -\ln\left[ 1-\frac{T_{13}}{15.9}\left( \frac{1}{e^{15.9/T_{ex}}-1}-0.0028\right)^{-1}\right],
        \end{equation}
        where $\tau_{13}$ denotes the optical depth of $^{13}$CO $(3\text{-}2)$ transition. Only those voxels that have an emission greater than 5$\sigma$ are considered real.

        \par The column density of $^{13}$CO $J=2$ level is calculated as:
        \begin{equation} \label{eq:LTE-5}
            \frac{N_{13} (J=2)}{\si{\per\square\cm}} = \frac{8\pi}{c^3} \frac{g_2}{g_3}\frac{\nu^3}{A_{32}}\frac{1}{1-e^{(-h\nu/k_BT_{ex})}} \, \tau_{\nu}\,dv ,
        \end{equation}
        where $g_2$ and $g_3$ are statistical weights of $J=2$ and $J=3$ rotational energy levels respectively; $A_{32} = 2.181\times10^{-6}\,\si{\per\s}$ is the Einstein's A coefficient for the $^{13}$CO $(3\text{-}2)$ transition \citep[taken from the Leiden Atomic and Molecular Database\footnote{\url{https://home.strw.leidenuniv.nl/~moldata/CO.html}};][]{sch}.
        \par The rotational partition function, $Z$, can be approximated by
        \begin{equation} \label{eq:LTE-6}
            Z \approx \frac{k_B}{hB} \left( T_{ex} + \frac{hB}{3k_B}\right),
        \end{equation}
        where $B=h/(8\pi^2\mathcal{I})$ is the rotation constant and is calculated using the moment of inertia $\mathcal{I}=\mu R_{CO}^2$, and $\mu$ is the reduced mass, and $R_{CO}=0.112\,\si{\nm}$ is the mean atomic separation of the CO molecule.
        \par Finally, using $N_{13}(J)$ and $Z$ one can calculate the total column density of $^{13}$CO using:
        \begin{equation} \label{eq:LTE-7}
            \frac{N_{13}}{\si{\per\square\cm}} = N_{13}(J) \frac{Z}{2J+1} \exp{\left[\frac{hBJ(J+1)}{k_BT_{ex}}\right]}
        \end{equation}
        
        \subsection*{Sub-Thermal Excitation}
        The analysis presented in this section is based on the assumption that LTE is applicable in all the voxels with detected emission. But for gas with densities below the critical density of CO (3\text{-}2) ($\approx 10^4\,\si{\per\cubic\cm}$), the energy level populations will not follow the Boltzmann distribution and consequently the measured excitation temperature will be lower than the actual temperature of the gas. An underestimation of the excitation temperature will lead to an overestimation of the column density values according to Eq.\ref{eq:LTE-5}. But, for optically thick emission the critical denisity can be effectively reduced (roughly by $1/\tau$) thereby making it possible for the gas to reach LTE at lower densities. We tried to test whether the limit of large optical depth for \ce{^{12}CO} holds in regions where \ce{^{13}CO} is also detected. For this, we took the lowest detected \ce{^{13}CO} column density at a $5\sigma$ significance and estimated the \ce{^{12}CO} column density adopting a \ce{^{12}CO}/\ce{^{13}CO} abundance ratio of 60.
        
        \begin{landscape}
            \begin{figure}[htpb]
                \includegraphics[trim = 0cm 0cm 0cm 0cm, clip,width=0.67\textwidth]{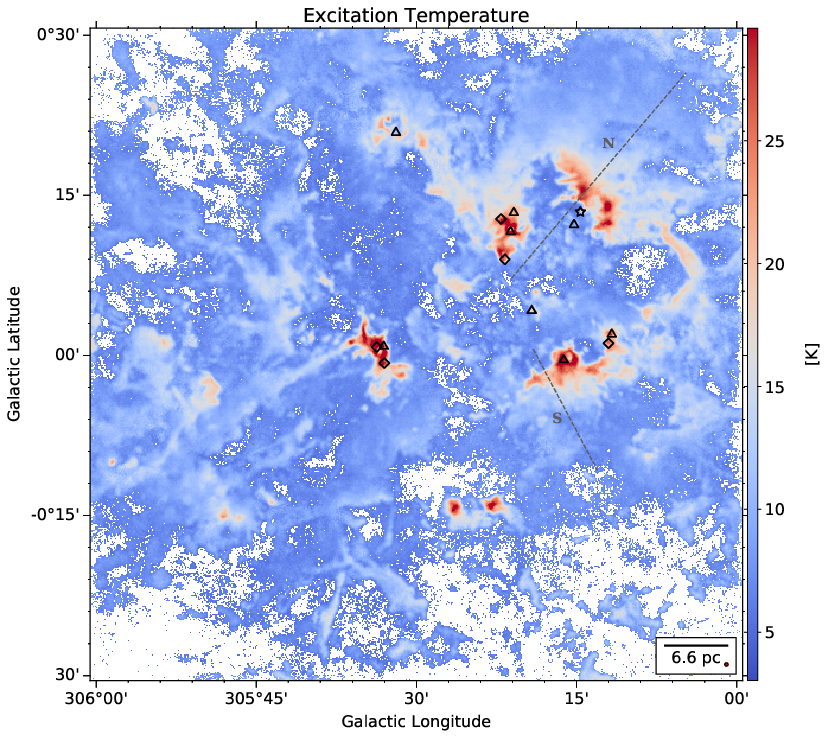}
                \includegraphics[trim = 0cm 0cm 0cm 0cm, clip,width=0.67\textwidth]{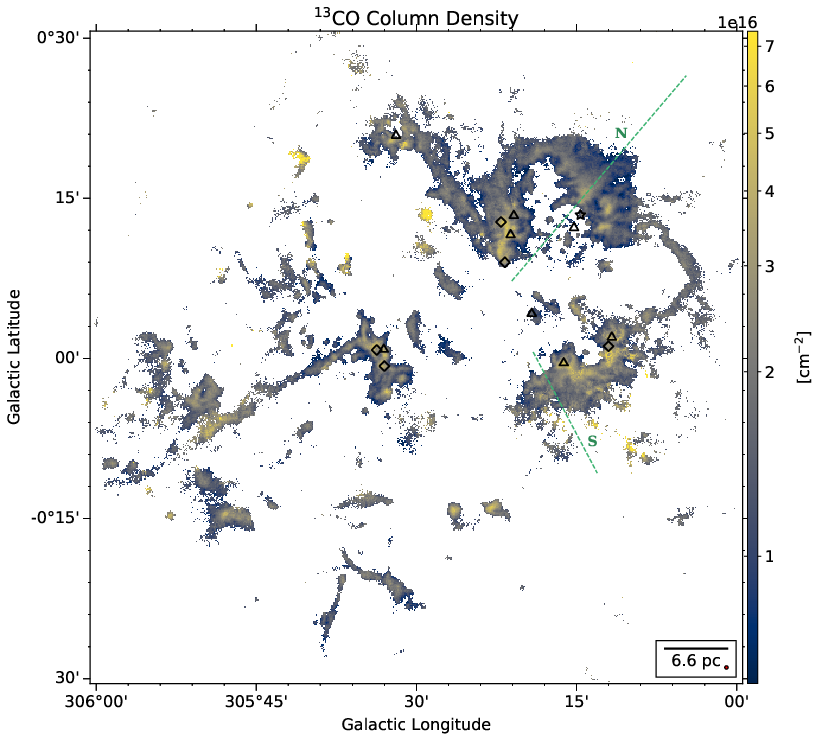}
                \caption{\textit{Left}: Mean $T_{ex}$ map of the G305 complex. For each pixel this was calculated by integrating over all the channels. Only those channels that had emission above 5$\sigma$ noise level were considered. \textit{Right} Integrated $^{13}$CO Column Density map of G305. This was calculated by summing the $^{13}$CO total column density per channel over all the channels. Overlaid on top, are the positions of the HII regions (triangles), UC HII regions (diamonds) and the BRC (star) as determined by \citet{hindson12,hindson13}.}
                \label{fig:integrated}
            \end{figure}
        \end{landscape}

        This value was calculated using the equation $12C/13C = 6.21D_{GC}+18.71$ from \citet{milam2005} where $D_{GC}\approx6.6\,\si{kpc}$ is the galactocentric distance of G305. The estimated \ce{^{12}CO} column density corresponding to the lowest detected \ce{^{13}CO} column density is then $\sim\,2.46\times10^{17}\,\si{\per\square\cm}$. Eq.\ref{eq:LTE-5} was then used to calculate the velocity integrated optical depth for \ce{^{12}CO}(3-2) transition. We obtained $\int \tau_{\nu}dv \approx 149.193\,\si{\km\per\s}$ for an excitation temperature of $15\,\si{\kelvin}$. Assuming a Gaussian line profile with a standard deviation of $\sim 8.5\,\si{\km\per\s}$ (see Sect.\ref{sec:ALP} for the justification of the chosen value for standard deviation), the optical depth at the centroid velocity is given by $\tau_{12} = \left( \int \tau_{\nu}dv\right) / \left(\sigma_v \cdot \sqrt{2\pi}\right) \approx 7$. So, for pixels with detected \ce{^{13}CO} emission, the \ce{^{12}CO} optical depth is greater than 1. This along with the findings of \citet{hindson10} that the volume density of clumps in G305 ranges between $10^3\text{--}10^4\,\si{\per\cubic\cm}$ means that the emission from the pixels where both \ce{^{12}CO} and \ce{^{13}CO} are detected are most likely in LTE. In the other regions where only \ce{^{12}CO} emission is detected, the condition of LTE may not always apply and should be kept in mind while interpreting the results derived from them.

\section{Effects of Feedback on the Molecular Gas \label{sec:feedback_effects}}

    \subsection{Excitation Temperature and Column Density maps \label{sec:integrated}}
        The integrated spectrum over the whole G305 region (see Fig. \ref{fig:spectra}) does not show any obvious sign of confusion along the line of sight (the emission appears to come from one local standard of rest (LSR) velocity without foreground confusion at any other velocity channel). Hence, properties like excitation temperature and column densities integrated over the whole linewidth are very likely to be a good representation of the corresponding physical properties of the GMC.
        
        Fig. \ref{fig:integrated} (\textit{left}) shows the excitation temperature map of the complex where each pixel represents the mean $^{12}$CO excitation temperature at the corresponding position. Fig. \ref{fig:integrated} (\textit{right}) shows the $^{13}$CO column density of the region summed over all velocity channels on a logarithmic scale. The hot regions towards the northern, southern and western molecular clouds are coincident with the HII and UC HII regions existing in the region \citep[see Fig. \ref{fig:integrated};][]{hindson12,hindson13}. The UC HII regions are also located in regions of high column density surrounded by a region with relatively lower column density. One also notes some unconnected small regions where the column density seems to be much higher than in some of the densest regions inside the larger molecular clouds, notably at approximately $(l,b) \sim (305\degree 42\arcmin,+0\degree 19\arcmin), (305\degree 30\arcmin,+0\degree 13\arcmin)$ and $(305\degree 10\arcmin,-0\degree 8\arcmin)$. Of these, only that at $\sim (305\degree 30\arcmin,+0\degree 13\arcmin)$ is associated with a diffuse H {\small {II}} region G305.4399+00.2103. For other sources we have not found any counterparts in the literature.
        \par
\begin{figure}
            \centering
            \includegraphics[trim = 0cm 3cm 0cm 2cm, clip,width=0.5\textwidth]{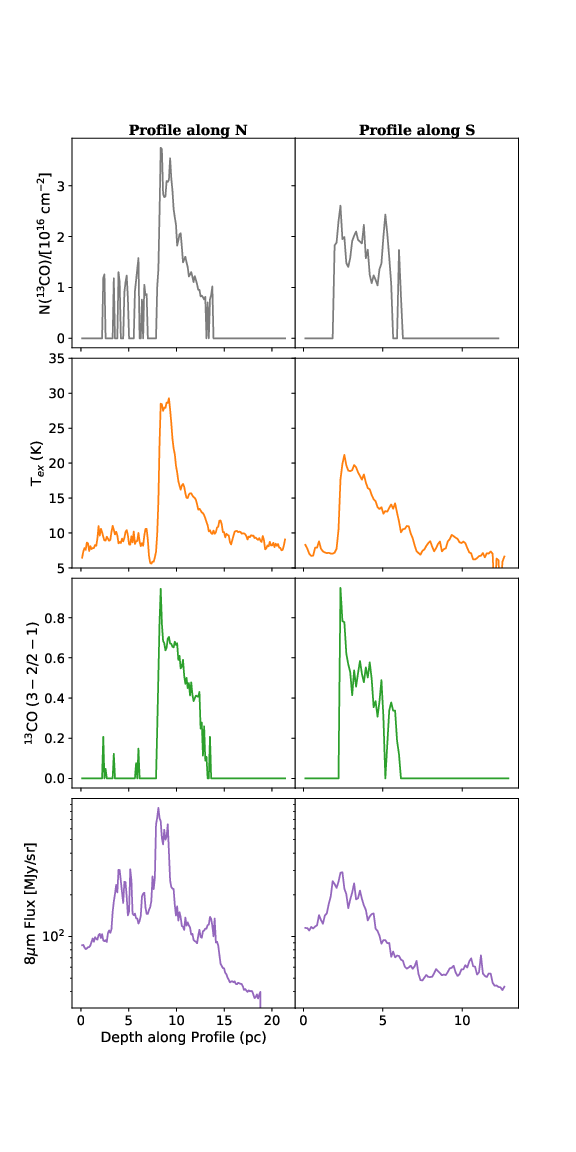}
            \caption{$^{13}$CO Column Density (grey), Excitation Temperature (orange), \ce{^{13}CO} $J=3\text{-}2$/$J=2\text{-}1$ ratio (green), and 8$\mu m$ Flux (purple) profiles for two separate directions in G305.  The profiles were plotted outwards from the center of the complex.}
            \label{fig:LTE_Profiles}
        \end{figure}

 In Fig.\ref{fig:integrated} we notice that the edge of the denser clouds show a marked jump in their excitation temperature at the side facing the central stellar clusters, which then tapers off as we move away from the center. This represents clear evidence that feedback from the stars acting on the edge of the dense gas is heating it. The exposed gas then acts as a shield for the rest of the cloud, resulting in a temperature profile that decreases as we move away from the center. Fig. \ref{fig:LTE_Profiles} shows the profiles of \ce{^{13}CO} column density (top panel row) and excitation temperature (second row of panels from the top) along two different arbitrarily selected directions leading away from the center as marked on Fig. \ref{fig:integrated}. These profiles have been obtained after smoothing the excitation temperature and column density maps to a resolution of 30$\si{\arcsecond}$ in order to compare them with other ancillary data as will be evident in the following sections. The left panel shows the profile along the line marked N and the right panel shows that along S. As is evident from the excitation temperature profiles, the gas is heated on the leading edge of the molecular cloud facing the central cavity. The feedback from the stars results in a steep increase in the temperature by a factor of $\sim 3$ at the edge facing the center. The temperature then steadily drops as we move away from the center. As pointed out in Sect.\ref{sec:lte} the excitation temperature before the leading edge of the profile may have been underestimated. So the increase in its value at the leading edge can be considered to be an upper limit. For the column density profiles, we notice that the northern profile also shows a pronounced evidence for compression as the column density at the leading edge is enhanced significantly compared to the trailing edge. However, for the southern profile even if the leading edge of the cloud has a higher column density than the rest of the cloud, the decrease in column density as we move away from the center is not as pronounced. Due to the absence of significant \ce{^{13}CO} emission between the stars and the dense gas boundary, we were only able to determine a lower limit on the extent of the region that has an increased column density at the edge of the cloud (based on the minimum column density ($\approx \num{4.1e15}\,\si{\per\square\cm}$ corresponding to $5\sigma$ times the rms noise value probed by our observations). This lower limit factor of $\gtrsim 5$ for S and $\gtrsim 10$ for N still represents a large increase in column density at the edge of the cloud.

    \subsection{Rotational excitation \label{sec:rot_excitation_studies}}
        \par The energy from the feedback being deposited onto the molecular cloud from the stars may be used up to excite the CO molecules to their higher $J$ rotational levels. In this section, we investigate whether feedback in G305 affects the rotational excitation in the molecular cloud. We complemented our data with $^{13}$CO $J=2\text{-}1$ data from the SEDIGISM \citep{sedigism} survey to create rotational excitation maps for the region.

        \par For this, the LAsMA $^{13}$CO $J=3\text{-}2$ map was first smoothed to the $30\arcsec$ angular resolution of the SEDIGISM map. In order to trace the excitation, the $^{13}$CO ~$J=3\text{-}2$/$J=2\text{-}1$ intensity ratio was calculated over the whole map. Pixels with a signal-to-noise ratio (S/N from now on) of less than 5 were blanked. Warm and dense gas is excited to higher $J$ levels resulting in a higher $^{13}$CO ($J=3\text{-}2$)/($J=2\text{-}1$) ratio.

        \begin{figure}
            \centering
            \includegraphics[width=0.5\textwidth]{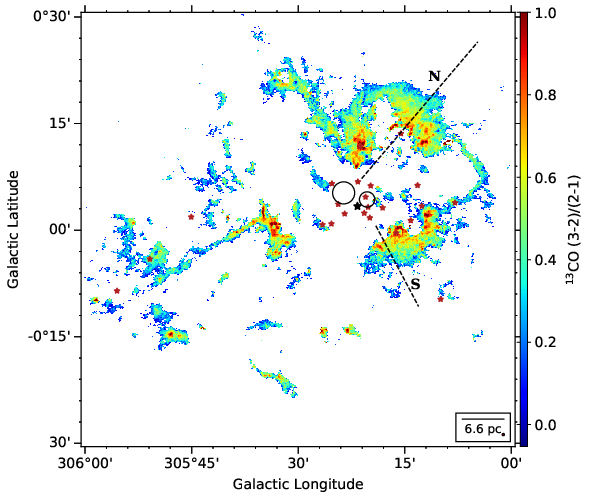}
            \caption{\ce{^{13}CO} $J=3\text{-}2$/$J=2\text{-}1$ ratio map of the G305 complex. The dashed lines show the direction along which the profiles of the excitation ratio was plotted in Fig. \ref{fig:LTE_Profiles}. The black circles show the locations of the Danks 1 and 2 clusters. The black star shows the position of WR48a, and the red stars show the positions of the stars from \citet{borissova}.}
            \label{fig:line_ratio_map}
        \end{figure}
        
        \par Fig. \ref{fig:line_ratio_map} shows the map of the ratio of the two line intensities. It is evident that this ratio is higher on the side of the cloud facing the central cavity. We also show two directional profiles cut through the excitation map in Fig.\ref{fig:LTE_Profiles} (third row of panels from the top). Both these profiles demonstrate the effect of feedback very well. In a very narrow region facing the center we see the excitation ratio almost equal to 1. For linear molecules, the rotational excitation can either increase by higher temperatures or larger volume densities or a combination of both, indicating a higher pressure at the cloud surfaces. As we then move away from the center, the excitation ratio decreases fast, indicating that the front end of the cloud is acting as a shield for the rest of the gas trailing it.

    \subsection{Energetics of Feedback}
\begin{figure}
            \includegraphics[width=0.5\textwidth]{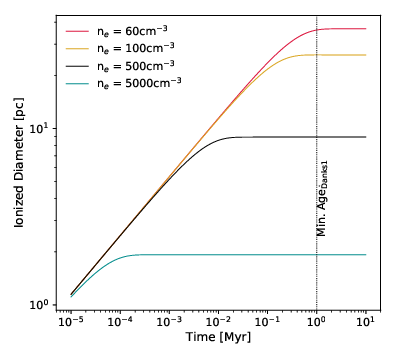}
            \caption{Time evolution of the ionization diameter of a spherical cloud of density $10^4\,\si{\per\cubic\cm}$ being illuminated by Danks1 and Danks2 star clusters at its center for different electron density values. The vertical dotted line is the minimum age of the Danks1 cluster take from \citet{davies12}.}
            \label{fig:ion_length}
        \end{figure}

        In this subsection, we try to explore whether the energy input from the central stars can account for the observed effects on the molecular gas morphology and excitation. Firstly, the OB stars inside the cavity of G305 can efficiently ionize the surrounding gas with the large amounts of ionizing photons they emit leading to photoevaporation of the cloud. Here, we estimate if this can lead to the observed size of the cavity given the age and type of stars in the complex. We follow the model used by \citet{watkins2019} for a simple case of a spherical cloud of uniform density $n_H$ being illuminated by a central stellar population emitting photons isotropically at rate $\mathcal{N}_i$. The radius of the central hole carved out by these photons in time $t$ given an electron density $n_e$ is given by the following equation.
        \begin{equation}
            R_{ion} = \left(\frac{3\mathcal{N}_i}{4 \pi \alpha_B n_e^2}[1-\text{exp}(-n_e^2 \alpha_B t/n_H)]\right)^{\frac{1}{3}},
        \end{equation}
        where $\alpha_B$ is the recombination coefficient. Considering the stars from the Danks1 and Danks2 as well as those outside the two clusters as reported in \citet{borissova} as the ionizing sources, we obtain $\mathcal{N}_i\approx 9.4\times10^{50}\,\si{\per\second}$ using the stellar classification from \citet{davies12,borissova} and calculating the ionizing flux for different stellar types using \citet{panagia1973} and \citet{crowther2007}. We adopt $n_H\approx 10^4\,\si{\per\cubic\cm}$ and $\alpha_B=2.7\times10^{-13}\,\si{\cubic\cm\per\s}$ \citep{hindson10,hindson13}, $n_e=[100-5000]\,\si{\per\cubic\cm}$. Figure \ref{fig:ion_length} shows the diameter of the ionized bubble as a function of time for a range of $n_e$ values. Given the diameter of the central cavity is $\sim\,10\,\text{pc}$ wide along the north-south direction (between regions V and VIII) and $\sim\,30\,\text{pc}$ across the east-west direction (between regions III and VII), this would imply that the population of visible stars inside the complex can drive such a cavity via photoionization given an $n_e\sim\, 10^2\,\si{\per\cubic\cm}$. But, this is a very simplistic view, as the starts in the cavity are not located at its center but are spread out over $\sim 20\,\si{pc}$. Also, electron density inside HII regions is not constant over time and varies from $10^2 \text{--} 10^5\,\si{\per\cubic\cm}$ depending on the diameter of the cavity as $n_e\propto D^{-1}$ \citep{kim2001,garay1999}. A more realistic scenario would be one where the expansion of the cavity is initially driven by the Danks 2 system for the first $1.5\,\si{Myr}$ which swepng out some of the ionized gas via strong winds (aided by the WR stars outside the Danks cluster) and thereby lowering the $n_e$ for the next generation of stars in Danks 1 and the others. The subsequent generations of stars could then clear out the gas more effectively.

        \par Next, we estimate whether the amount of energy input from the stars is large enough to cause the observed trend in the excitation line profiles, even if the nature of the profiles strongly suggest that feedback from the stars is heating the gas. For this we first calculate the external pressure exterted by the stars on the cloud surface. Assuming that most of the energy input from stellar radiation comes from the ionizing photons, the pressure from stars can be estimated as $P_{\rm{star}} = \mathcal{N}_i \, \langle h\nu \rangle / 4\pi {R_s}^2 \, c \, k$, where $\langle h\nu \rangle$ is the mean photon energy of an O-star (assumed to be $\sim15\,\rm{eV}$ \citep{Pellegrini07}), $R_s$ is the distance of the cloud from the emitting star, $c$ is the speed of light and $k$ is the Boltzmann's constant. We assume that the O5-O6V, B0-B1V and B2-B3V stars found by \citet{leistra2005} in the G305.3+0.2 cluster are the stars responsible for the observed excitation profile N and Danks1 is responsible for that in S. But the spectral classification of stars in Danks1 by \citet{davies12} spans a large range of ionising flux values. In order to obtain a lower limit on the ionising flux, we used the spectral type with the lowest ionising flux, i.e. O6V for O4-6, O8V for O6-8/8If, B3V for O8-B3, B3I for O8-B3I and WN9 for WNLh. We estimate a distance of 2.5\,pc between the edge of the profile N and the stars and $\sim 4$\,pc between Danks1 and the edge of S. Using these values we obtain the external pressure from the stars for N and S to be $P_{\rm{star, N}} \sim 2.5\times 10^5\,\si{\kelvin\per\cubic\cm}$, and $P_{\rm{star, S}} \sim 2.6\times 10^5\,\si{\kelvin\per\cubic\cm}$ respectively. We then calculate the thermal pressure at the edge of the cloud facing the cavity and that at the far end away from the cavity. The pressure exerted by thermal motion inside a cloud can be estimated as $P_{th}=n(H_2)\,T\,\si{\kelvin\per\cubic\cm}$. For the far end of the cloud in both cases N and S, we assume the density to be equal to the average value of clump densities from Table 5 in \citet{hindson13}, i.e. $n(H_2)\sim 1.5\times10^3\,\si{\per\cubic\cm}$. For the temperature we adopt a value of $T\sim10\,\si{\kelvin}$ for both N and S (see Fig.\ref{fig:LTE_Profiles}). So, for both N and S the thermal pressure in the far end of the cloud is $P^{far}_{th} = 1.5\times10^4\,\si{\kelvin\per\cubic\cm}$. For the side of the profiles N and S facing the cavity, we need to deduce the value of $n_{H_2}$. For this we assume that the pressure from the stars only compresses the cloud in a direction along the plane of the sky. We can then equate the proportional increase in column density to that of the observed volume density. From Fig.\ref{fig:LTE_Profiles} we calculate the proportional increase in column density at the edge facing the center compared to that away from the cavity to be $\sim 4$ for N and $\sim 1.6$ for S. Now calculating the thermal pressure at the edge of the profiles facing the central cavity we obtain $P^{near}_{th,N}=28\cdot4\cdot1.5\times10^3 = 1.68\times10^5\,\si{\kelvin\per\cubic\cm}$ and  $P^{near}_{th,S}=22\cdot1.6\cdot1.5\times10^3 = 5.28\times10^4\,\si{\kelvin\per\cubic\cm}$. The difference in the thermal pressure between the leading and trailing edge for N and S is therefore $\Delta P_{th, N} \sim 1.53\times10^5\,\si{\kelvin\per\cubic\cm}$ and $\Delta P_{th, S} \sim 3.78\times10^4\,\si{\kelvin\per\cubic\cm}$ respectively. This difference in thermal pressure accounts for $\sim 61\%$ of $P_{\rm{star, N}}$ and less than $\sim 14.5\%$ of $P_{\rm{star, S}}$. Hence, for the profile N about $61\%$ of the radiation pressure is going towards heating up the gas whereas for S this is less than $15\%$.
        
\section{Characterizing Feedback in G305} \label{sec:feedback-overall}
 
    \par So far, the evidence of feedback on the molecular gas in G305 has been mostly based on morphology, with specific examples of profiles along selected directions. In this section we attempt to quantitatively study the effect of feedback on the \textit{excitation} of the gas at a global level over the whole cloud complex. In order to see this global nature of the gas excitation in feedback regions, we first need to identify as well as quantify where these feedback regions are.
    
    \subsection{Identifying feedback regions : GLIMPSE 8$\mu \rm{m}$ map}
        \par The 8$\,\si{\micro\metre}$ map obtained from the Galactic Legacy Infrared Mid-Plane Survey Extraordinaire \citep[GLIMPSE:][]{Benjamin_glimpse,Churchwell_glimpse} is a very useful tool to identify the regions of stellar feedback.

        \begin{figure}[h]
            \centering
            \includegraphics[width=0.5\textwidth]{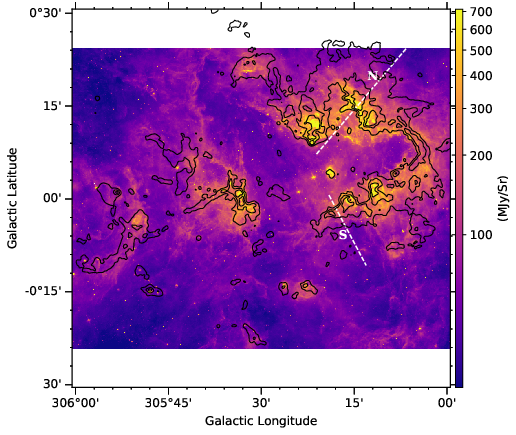}
            \caption{GLIMPSE $8\,\si{\micro\metre}$ map of the G305 regions. The black contours correspond to the $^{12}$CO J=3--2 integrated intensities of 30, 70 and 120$\,\si{\kelvin\km\per\second}$ (50, 120 and 200$\sigma$, respectively). The dashed lines show the direction along which the profiles of the excitation ratio was plotted in Fig. \ref{fig:LTE_Profiles}.}
            \label{fig:8mu}
        \end{figure}

        \par PDRs mark the boundary between the ionized and neutral gas in a molecular cloud \citep{rathborne}. The FUV photons from stars excite polycyclic aromatic hydrocarbons (PAHs) on the surface of dense molecular clouds at the interface between the ionization front and the molecular gas \citep{Tielens2008}. The PAHs absorb FUV radiation from hot stars and re-emit fluorescently in several broad bands at near and mid infrared (IR) wavelengths \citep{Allamandola1989}. The most luminous of these is within the $6.5 \text{ to }9\,\si{\micro\metre}$ range covered by the 8$\,\mu\si{\metre}$ filter of the Infrared Array Camera \citep[IRAC:][]{Fazio2004} aboard the Spitzer Space Observatory that was employed for GLIMPSE. Hence, bright regions in 8$\,\si{\micro\metre}$ Spitzer/IRAC images correspond to regions subjected to large amounts of radiative feedback to FUV radiation from the stars.

        \par Fig. \ref{fig:8mu} shows the Spitzer 8$\,\mu\si{\metre}$ image of the G305 region. The molecular gas in the region seems to be coextensive with the bright 8$\,\mu\si{\metre}$ emission. All of the CO emission in the region is associated with IR emission. In order to demonstrate the validity of the 8$\,\mu\si{\metre}$ flux as a tracer for feedback we plotted its profiles along the same direction as the excitation temperature, column density and the $^{13}$CO $J=3\text{-}2$/$J=2\text{-}1$ intensity ratio in Fig. \ref{fig:LTE_Profiles}. Before plotting the profile, the 8$\,\si{\micro\metre}$ map was first smoothed to the same resolution as the other excitation maps of 30$\si{\arcsecond}$. At the edges of the higher column density clouds, we observe a sharp increase in the 8$\,\mu\si{\metre}$ flux (around a depth of $7\,\si{pc}$ for N and $1.5\,\si{pc}$ for S), marking the brightest parts of the PDRs of the clouds. This increase coincides with the increase in the gas excitation properties. Afterwards, all three properties decrease along with the 8$\,\mu\si{\metre}$ flux. The decreasing profile as we move away from the center also demonstrates that the PDR is effectively shielding the molecular cloud, absorbing the bulk of the FUV photons. Inside the denser cloud along N, the 8$\,\mu\si{\metre}$ profile also shows a second local peak (around a depth of $13\,\si{pc}$ along the profile and at a depth of $\sim\,6\,\si{pc}$ from the edge of the high density cloud) before sharply falling of. This also coincides with the local peak in the column density as well as the rotational excitation (third panel in Fig.\ref{fig:LTE_Profiles}) profiles. But, the excitation temperature (first panel) does not show a distinct peak. An additional peak also exists at a depth of about $4\,\si{pc}$ along N. This along with the peak coincident with the edge of the high density cloud appears to be part of a bubble shaped structure in Fig.\ref{fig:8mu}. There is very little dense molecular material left around the former peak and the diffuse gas does not appear to be impacted from the feedback as can be seen from the excitation temperature profile corresponding to this peak. Some pockets of higher column density gas appear to be coincident with this peak. The feedback from the stars reported in \citet{leistra2005} which are located inside the bubble are most likely responsible for this structure.

        \par This 8\,$\mu \si{\metre}$ map was used as a template to identify regions of feedback based on their integrated flux. In order to quantify the feedback itself we assumed the 8$\,\si{\micro\metre}$ intensity to be a proxy for the feedback by radiation, i.e. higher intensity corresponds to a stronger feedback. The properties of the molecular CO emission were then investigated in these feedback regions.
    
    \subsection{Molecular Gas Properties vs 8\,$\mu$m Flux} \label{sec:excitation_studies}

        \par In this section we investigate how the excitation temperature derived from \ce{^{12}CO} emission, the \ce{^{12}CO} 3--2/2--1 intensity ratio, and the \ce{^{13}CO} column density vary with the degree of feedback traced by the integrated 8$\mu$m flux. For this, the regions within a given interval of $8~\si{\micro\metre}$ integrated flux were masked.
        
        \par
\begin{figure*}
          \includegraphics[width=0.328\textwidth]{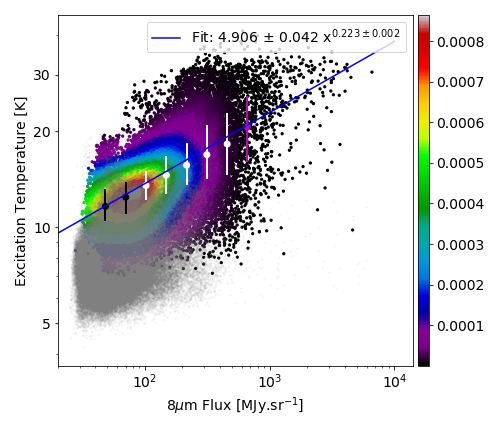}
          \includegraphics[width=0.33\textwidth]{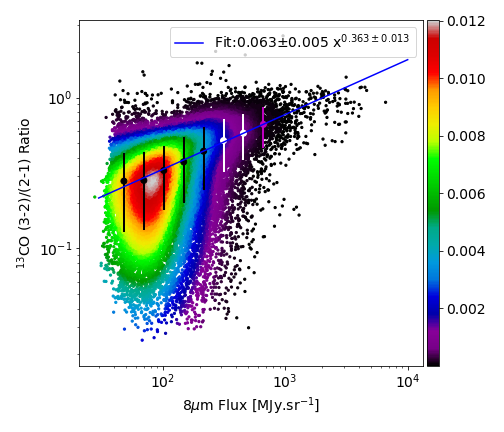}
          \includegraphics[width=0.34\textwidth]{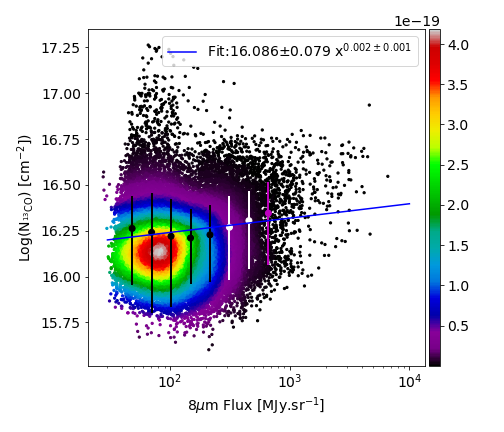}
          \caption{A pixel by pixel scatter plot of different gas properties versus $8\,\mu$m flux. The colors represents the probability density of the scatter points obtained by a kernel density estimate using Gaussian kernels. Overlaid on top are the mean values of the quantity probed along with their standard deviations plotted as a function of 8$\,\mu$m flux. The blue line shows the power law fit to the mean values vs 8$\,\mu$m flux. The results of the fit are shown at the top right corner of each panel. The gray translucent scatter points in the excitation temperature plot (\textit{left}) correspond to the pixels which do not have a corresponding \ce{^{13}CO} detection and are not included in the power law fit.}
          \label{fig:scatterplots}
        \end{figure*}

 For every 8$\,\mu\si{\metre}$ flux interval, we investigated the aforementioned properties of the molecular emission of pixels within that mask. Since the GLIMPSE image does not cover the whole range of latitude of the LAsMA map, only the overlapping range of latitude was considered for the analysis. Moreover, the blanked pixels in the excitation map were ignored. A pixel scatter plot was then made for the 8$\,\mu$m flux versus each of the three properties (see Fig.\ref{fig:scatterplots}). Only those pixels with S/N>5 for all the properties being investigated were considered. This also ensured that the pixels with possible sub-thermal emission were avoided (these pixels are shown as translucent gray scatter plots in the figure). The color of the scatter points show the density of points in its vicinity. The mean values of the properties are plotted as a function of 8$\,\mu$m flux on top of the pixel scatter plot. A power-law function was then fitted to the mean values for all the three properties and the results of the fits are also shown in the figure. As is evident from Fig.\ref{fig:scatterplots} excitation and rotational temperatures increase with increasing 8$\,\mu$m flux. However, we only see a very modest increase in the median column density values for higher 8$\,\mu\si{\metre}$ flux. Additionally, we also see a set of scatter points which have a much steeper slope between 50 and 150$\,$MJy.sr$^{-1}$ for column density. This appears to be mostly from the regions which are far away from the central stars and receive very little feedback from them. Only one of these regions i.e. G305.4399+00.2103 is known to be a diffuse H II region in literature \citep{urquhart2014rms}. The other regions could also be pre-stellar sources collapsing under gravitation and therefore show high column densities. It is beyond the scope of this paper to explore the reason for the observed trend at lower 8$\,\mu$m flux.
        
\section{Dynamics of gas under feedback \label{sec:dynamics}}

    \par In the preceding sections we have investigated the effects of feedback on the morphology of the gas and its excitation. In this section we will investigate the dynamical signature of this feedback. The shape of the spectrum in Fig. \ref{fig:spectra} is the sum of the spectra over all the pixels in the region. As has been seen already, the G305 region consists of various clouds moving at different velocities with respect to us. Additionally, the shape of the profile varies from pixel to pixel. It is impossible to disentangle the contribution of the line centroid velocities and the shape of the line profile from each pixel on the overall shape of the profile in Fig. \ref{fig:spectra}. In the following sections we will try to study these line characteristics in a statistical way.
    
    \subsection{Velocity Centroid Probability Distribution Function} \label{sec:v-pdf}

        \par The probability distribution function (PDF) of velocities in observational or simulated datasets can be used to characterize a cloud's velocity structure. PDFs can show the degree of intermittency in the turbulent molecular cloud \citep{falgarone-philips-1990} through the shapes of their wings. Increasing intermittency causes a transition from Gaussian to exponential wings in velocity PDFs. In this section, we estimate the velocity PDF of G305 in order to reveal further effects of feedback on the molecular gas.
        
        \par It is not possible to know the true velocity PDFs of the clouds from their observations, owing to the fact that the complete velocity information is not available for all three dimensions. Historically, two methods have been employed to deduce the true velocity PDF from observational data of spectral lines: \citet{kleiner-diskman-1985,miesch-bally-1994,miesch-1999,ossenkopf-low-2002} used the distribution of the line centroid velocities to deduce the velocity PDFs. Alternatively, \citet{falgarone-philips-1990} estimated the velocity PDFs from high S/N observations of single line profiles. \citet{ossenkopf-low-2002} tested the realm of validity of both of these approaches and concluded that if the size of the observed map is larger or comparable to the depth of the cloud, the centroid velocity PDF reproduces the correct velocity distribution. In contrast, for small maps, the average line profiles comprise of a more comprehensive sampling of velocities, given their more comprehensive line-of-sight sampling and hence, provide a better approximation to the true velocity structure of the cloud.

        \par The G305 complex is believed to have a flattened geometry \citep{hindson13} in the plane of the sky. Hence, instead of the average line profile shown in Fig. \ref{fig:spectra}, the centroid velocity PDF of the complex should be a good tracer of the actual velocity distribution. An added benefit of using the velocity centroid PDFs over using average line profiles is that local phenomena such as outflows that could bias the determination of the global velocity structure do not affect velocity centroid PDFs, since the broad wings of their line profiles leave the centroid velocity unaffected.
        
        \par The centroid velocity is effectively the moment-1 value of each pixel. Its values were calculated from the \ce{^{12}CO} line using the formula,
        \begin{equation} \label{eq:1}
            \frac{v_{c}}{\si{\km\per\second}} = \frac{\sum_{i=1}^{N_{chan}} T_i \cdot v_i}{\sum_{i=1}^{N_{chan}}T_i}
        \end{equation}
        where, $N_{chan}$ is the total number of channels and $T_i$ is the observed antenna temperature of the corresponding channel. Similarly to the moment-0 maps, only channels with signal greater than $5\sigma$ were included in the calculation. Once the centroid velocity for each pixel was obtained, the PDF was estimated using a normal histogram (i.e. the sum of the PDF $P$ is normalized to unity). Pixels were partially weighted based on their S/N. Pixels with a S/N greater than 100 were assigned an S/N of 100 in order to prevent pixels with significant emission, with S/N $>$, say, 10 (but $<< 100$) to be unreasonably downweighted.

        \begin{figure}
            \centering
            \includegraphics[width=0.4\textwidth]{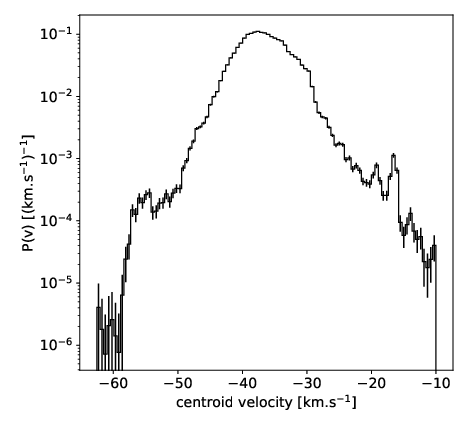}
            \caption{Probability density distribution of centroid velocities of \ce{^{12}CO} for the G305 molecular cloud complex. The error bars are statistical errors calculated as the square root of the histogram amplitudes.}
            \label{fig:v_centroid_pdf}
        \end{figure}

        \par Fig. \ref{fig:v_centroid_pdf} shows the centroid velocity PDF for the G305 complex. The shape of the wings in the velocity PDF suggest that the velocity distribution in the complex is not consistent with a Gaussian which would appear as a parabola on a log-lin plot. To quantify the shape of the distribution we calculate its statistical moments. The most frequently used moments are the mean ($\langle v_c \rangle$), variance ($\sigma^2$), and the Kurtosis ($K$) of the distribution defined as below:

        \begin{align}
            & \frac{\langle v_c \rangle}{\si{\km\per\second}} = \int_{-\infty}^{+\infty} dv_c P(v_c) v_c \label{eq:2} \\
            & \frac{\sigma^2}{\si{\square\km\per\square\second}} = \int_{-\infty}^{+\infty} dv_c P(v_c) \lbrack v_c - \langle v_c \rangle \rbrack ^2 \label{eq:3} \\
            & K = \frac{1}{\sigma^4} \int_{-\infty}^{+\infty} dv_c P(v_c) \lbrack v_c - \langle v_c \rangle \rbrack ^4 \label{eq:4}.
        \end{align}
        The variance is a measure of the total turbulent mixing energy. The Kurtosis is a measure of the deviation from a Gaussian distribution. It assumes a value of three for a Gaussian distribution and six for a distribution with exponential wings. For our PDF we obtain the following values for the aforementioned moments :

        \begin{align*}
            & \langle v_c \rangle = -37 \pm 7 \, \si{\km\per\s} \\
            & \sigma^2 = 4.0 \pm 0.9 \, \si{\square\km\per\square\second} \\
            & K = 5.1 \pm 1.0
        \end{align*}
        
        \par The Kurtosis of $5.1 \pm 1.0$ indicates that the velocity PDF of the complex has exponential wings. 
        Using two-dimensional Burgers turbulence simulations, neglecting pressure forces, \citet{chappell-scalo-2001} showed that the velocity PDFs are Gaussian for models of decaying turbulence and have exponential wings for models driven by strong stellar winds.
        
    \subsection{Stacked Spectra} \label{sec:ALP}

        \begin{figure*}
            \centering
            \includegraphics[width=\textwidth]{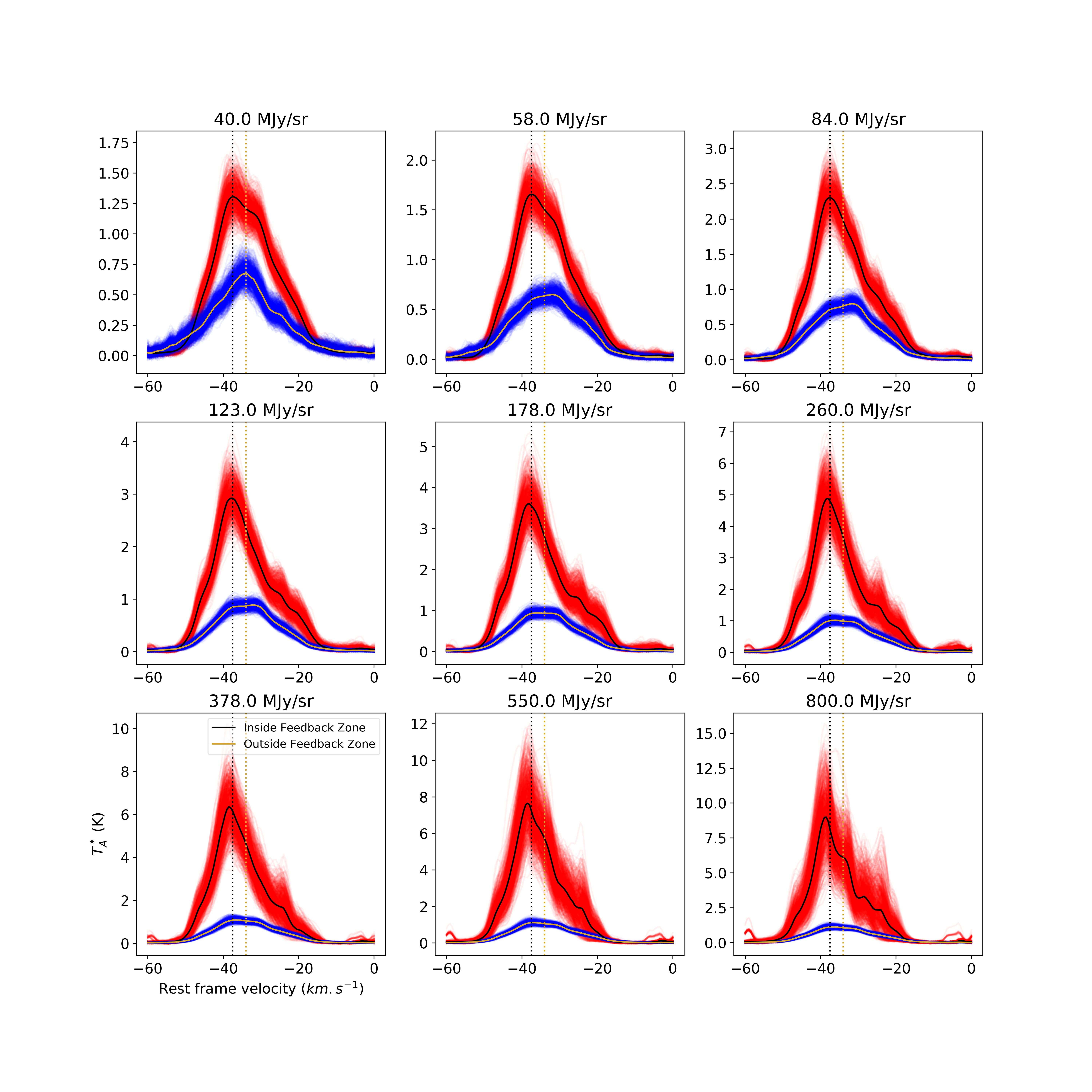}
            \caption{Stacked spectra of \ce{^{12}CO}(J=3--2) emission corresponding to the \textit{inside-} (translucent red) and \textit{outside feedback zones} (translucent blue) for different choices of GLIMPSE 8$\,\mu\si{\metre}$ flux thresholds (specified on top of each sub-plot). Each sub-plot consists of 500 stacked spectra each for both zones obtained by randomly selecting the square root of the total number of pixels from each zone. The black and golden spectra for each zone were obtained by calculating the median value for each channel over the 500 iterations performed. The black (golden) dotted line in all the sub-plots correspond to the position of peak emission of the median stacked spectra inside (outside) the feedback zone corresponding to the threshold of $40\,\si{\mega\jansky\per\steradian}$.}
            \label{fig:ALP-spectra-threshold}
        \end{figure*}

        \begin{figure*}
            \centering
            \includegraphics[width=\textwidth]{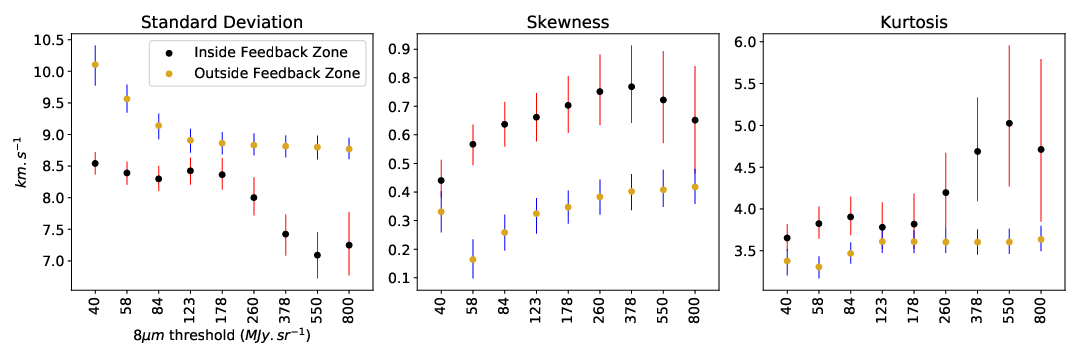}
            \caption{Variance (left panel), skewness (middle panel) and Kurtosis (right panel) of stacked spectra corresponding to \textit{inside-} (black dots with red bars) and \textit{outside feedback zones} (golden dots with blue bars) in Fig. \ref{fig:ALP-spectra-threshold} as a function of GLIMPSE 8$\,\mu\si{\metre}$ flux thresholds. The dots represent the median value and the bars span the 25 to 75 percentile spread of values from the 500 iterations for each threshold value.}
            \label{fig:ALP-statistics-threshold}
        \end{figure*}

        \par The velocity centroid PDFs as seen in Sect.~\ref{sec:v-pdf} contain information only about the central velocity of the gas. In this section we investigate the effect of feedback on the line profiles of the G305 molecular cloud complex, especially whether the shape of the line profile in the regions where we expect feedback to be present is significantly different from those where we see little evidence of feedback. We also try to characterize the deviations, if any, as a function of the strength of the feedback.

        \par We used the same method as in section \ref{sec:excitation_studies} to quantify the strength of the feedback. Different 8$\,\mu\si{\metre}$ intensities were logarithmically distributed. These values were used as a threshold to create masks on the G305 region. Regions with 8$\,\mu\si{\metre}$ flux greater than the threshold were labeled as \textit{inside feedback zone} and those outside were labeled \textit{outside feedback zone}. All the maps were reprojected onto the same two dimensional grid over the galactic longitude and latitude. All the spectra were then translated to a common central velocity based on their velocity centroids (see section \ref{sec:v-pdf} for how to calculate the velocity centroids). After aligning, a number of pixels equal to the square root of the total pixels in the respective zones were randomly selected. For these randomly chosen pixels the spectra were averaged to obtain an average spectrum representing the \textit{inside-} and \textit{outside feedback zones}. The spectra obtained by this method will be referred to as average stacked spectrum from now on. This process was then repeated 500 times to avoid any biases and obtain a more representative set of spectra for each region.

        \par  Fig. \ref{fig:ALP-spectra-threshold} shows the stacked spectra corresponding to the \textit{inside-} and \textit{outside feedback zones} for the G305 molecular cloud complex. We also derived a median stacked spectrum for each zone corresponding to all thresholds. This was obtained by calculating the median value for each channel over the 500 iterations performed. A few things become evident from Fig. \ref{fig:ALP-spectra-threshold}. The spectra from the \textit{inside feedback zones} are brighter than \textit{outside feedback zones}. With increasing feedback strength the peaks of the former also get brighter indicating that with stronger feedback more gas gets excited to the higher $J$ transitions. There is not much overlap between the individual stacked spectra from the \textit{outside-} and \textit{inside feedback zones}. Hence, the median profiles for the two regions for each threshold which look very different from each other are indeed representative of the actual differences between the two regions and are not in fact biased by any individual stacked spectra. This was also confirmed by running a Kolmogorov-Smirnov (KS) test \citep{ks-test-1,ks-test-2} between the spectra inside and outside the thresholds. We placed $p<0.05$ constraint to reject the null hypothesis that the spectra are drawn from the same population. The p-value measures the probability of random chance being responsible for the observed difference between the two spectra. The test rejected the null hypothesis with overwhelming certainty ($p<10^{-5}$). The median stacked spectrum from the \textit{inside feedback zones} appear to be blue shifted compared to that from the \textit{outside feedback zones}. This is an consequence of the translation of the spectra to a common velocity using their moment-1 values. The spectra inside the feedback zones have a peak component that is at the same velocity as that outside the feedback zone. But in addition the feedback from the stars seems to have pushed some gas away from us which shows up as the broad positively skewed part of the spectrum inside the feedback zone. This makes the moment-1 value of the overall spectrum to be redshifted from the true peak of the spectrum. Consequently, in the process of aligning all spectra along their moment-1 value, those inside the feedback zone appear to be blue shifted. Thus, the apparent blue shift of the stacked spectra from inside the feedback zone is evidence of the stellar feedback pushing gas away from us.
        
        \par In order to characterize the stacked spectra in Fig. \ref{fig:ALP-spectra-threshold} we calculated their statistical moments. The moments used for these profiles are the variance ($\sigma^2)$, skewness (S), and Kurtosis (K) :
        \begin{align}
            & \frac{\sigma^2}{\si{\square\km\per\square\second}} = \left(\int_{-\infty}^{+\infty} dv P(v) \lbrack v - \langle v \rangle \rbrack ^2\right)/\left(\int_{-\infty}^{+\infty} dv P(v)\right) \label{eq:5} \\
            & S = \frac{1}{\sigma^3} \cdot \left(\int_{-\infty}^{+\infty} dv P(v) \lbrack v - \langle v \rangle \rbrack ^3\right)/\left(\int_{-\infty}^{+\infty} dv P(v)\right) \label{eq:6} \\
            & K = \frac{1}{\sigma^4} \cdot \left(\int_{-\infty}^{+\infty} dv P(v) \lbrack v - \langle v \rangle \rbrack ^4\right)/\left(\int_{-\infty}^{+\infty} dv P(v)\right) \label{eq:7}
        \end{align}
        Here, $P(v)$ is the antenna temperature of the spectrum at the velocity $v$, and $\langle v \rangle$ is the expectation value for the velocity of the spectrum given by $\langle v \rangle = \left(\int_{-\infty}^{+\infty} dv P(v) v \right)/\left(\int_{-\infty}^{+\infty} dv P(v)\right)\,\si{\km\per\second}$.
        
        \par Fig. \ref{fig:ALP-statistics-threshold} demonstrates how these moments depend on the choice of the 8$\,\mu\si{\metre}$ flux threshold. A few things stand out from this result. We expect turbulence to result in broader line profiles. But contrary to expectations, the variance of the average stacked spectra from the \textit{inside feedback zones} are mostly smaller than that from the \textit{outside feedback zone}. For the \textit{outside feedback zones} the variance decreases until $\sim 80 \, \si{\mega\jansky\per\steradian}$ and then stays constant with increasing threshold. Looking at the variances of average stacked spectra for the \textit{inside feedback zones}, we observe that the variance stays constant irrespective of the choice of threshold until $\sim 180 \, \si{\mega\jansky\per\steradian}$ when it starts decreasing. We observe that the stacked spectra are consistently more skewed inside the feedback regions than outside irrespective of the choice of threshold. A positive skewness implies that the profile's red wing is enhanced compared to the blue wing. As explained before, this hints at the gas being pushed away from us. For small threshold values, the stacked spectra in \textit{outside feedback zones} have skewness close to zero. But, it increases with increasing threshold levels. Kurtosis of the stacked spectra in the \textit{inside feedback zones} are mostly larger than those in the \textit{outside feedback zone} indicating that the profiles have more pronounced wings for the gas impacted with feedback. For 8$\,\mu\si{\metre}$ flux thresholds above $\sim 180 \, \si{\mega\jansky\per\steradian}$, the Kurtosis of the spectra inside the feedback zone increases. For very high threshold values the Kurtosis shows a large spread. These regions correspond to those very close to the UC HII regions and hence show that the feedback in these regions is very effective at pushing gas away. This is also evident from the stacked spectra in Fig.\ref{fig:ALP-spectra-threshold} where secondary peaks are clearly visible showing gas clumps being pushed out. The Kurtosis of the spectra outside the feedback zone almost stays in the same range. We also observed that the trend followed by the Kurtosis is exactly opposite to the one followed by the variances for both inside and outside feedback zones. When the variances of stacked spectra decrease, the Kurtosis increases and vice versa.

        \subsection{Stacked Spectra for 8$\,\mu\si{\metre}$ Flux Intervals} \label{sec:ALP-interval}

            \begin{figure*}
             \centering
             \includegraphics[width=\textwidth]{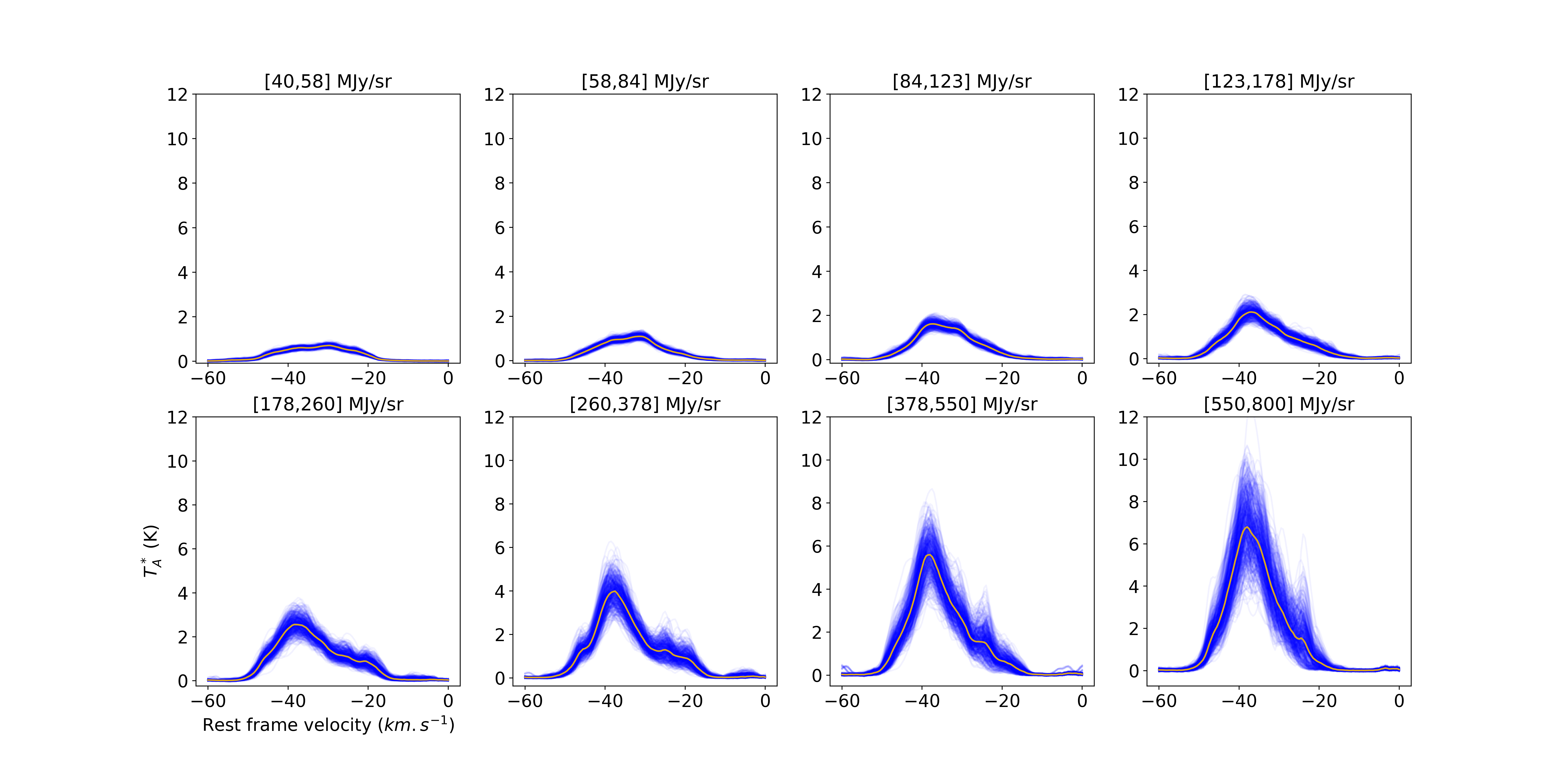}
             \caption{Stacked Spectra for regions corresponding to different choices of GLIMPSE 8$\,\mu\si{\metre}$ flux intervals. Each interval sub-plot consists of 500 stacked spectra (translucent blue), each corresponding to randomly selected pixels from the region. The golden spectra in each sub-plot corresponds to the median spectra for that interval obtained by taking the median channel intensity for each channel. The interval values are given on top of each sub-plot.}
             \label{fig:ALP_spectra_intervals}
            \end{figure*}

            \begin{figure*}
             \centering
             \includegraphics[width=\textwidth]{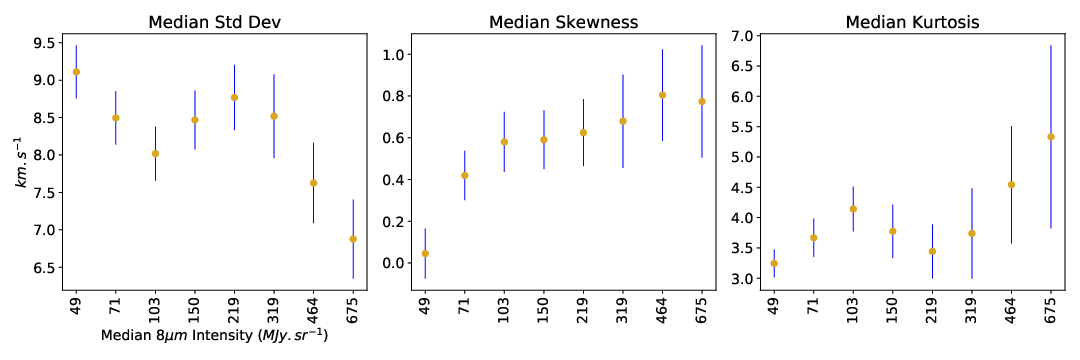}
             \caption{Variance (left panel), skewness (middle panel) and Kurtosis (right panel) of stacked spectra corresponding to GLIMPSE 8$\,\mu\si{\metre}$ flux intervals in Fig. \ref{fig:ALP_spectra_intervals} as a function of the median 8$\,\mu\si{\metre}$ flux for that interval. The golden circles represent the median of the distribution and the blue lines span the 5 to 95 percentile range of values for each set of 500 iterations per interval.}
             \label{fig:ALP_statistic_intervals}
            \end{figure*}

            \par In the preceding section, the stacked spectra inside the feedback zone for a given threshold include the cumulative sum of the contributions from all pixels with 8$\,\mu\si{\metre}$ flux greater than the threshold value. It becomes difficult to identify the spectra representative of a given interval of 8$\,\mu\si{\metre}$ flux. For this, we decided to investigate the stacked spectra corresponding to intervals of 8$\,\mu\si{\metre}$ flux. Pixels corresponding to the flux intervals were masked and their stacked spectra were derived by randomly picking pixels corresponding to each mask, similar to that done in Sec.\ref{sec:ALP}. The process was repeated 500 times. A median spectrum was derived from the set of 500 stacked spectra corresponding to each mask by taking the median intensity of each channel.

            Fig. \ref{fig:ALP_spectra_intervals} shows the stacked spectra corresponding to the pixels in 8$\,\mu\si{\metre}$ flux intervals. The peaks of the profiles do increase with increasing flux as expected. In order to look at their properties we calculated the statistical moments of each of the stacked spectra using Eq.\ref{eq:5} -- \ref{eq:7}. Their median along with the 5 to 95 percentile range of values spanned by the spread were then plotted as a function of the median value of the 8$\,\mu\si{\metre}$ flux interval. Fig. \ref{fig:ALP_statistic_intervals} shows the statistical moments as a function of 8$\,\mu\si{\metre}$ flux. The standard deviation initially decreases with increasing 8$\,\mu\si{\metre}$ flux but starts increasing with flux from $\sim 100 \, \si{\mega\jansky\per\steradian}$ up until $\sim 215 \, \si{\mega\jansky\per\steradian}$. After this the standard deviation of the stacked spectra decreases. The Kurtosis of the stacked spectra show the exact opposite trends to the standard deviation: where if the standard deviation decreases, the Kurtosis increases and vice versa. At the lowest range of probed 8$\,\mu\si{\metre}$ flux the stacked spectra seem to have a skewness approaching 0 and a Kurtosis close to 3, which hints at a Gaussian shape of its profile. The skewness of the stacked spectra then increases to 0.6 for an 8$\,\mu\si{\metre}$ flux of $\sim 100 \, \si{\mega\jansky\per\steradian}$, while it increases only a little (to 0.8) for higher values, around $\sim 600 \, \si{\mega\jansky\per\steradian}$.
            
            \par
\begin{figure}
            \centering
            \includegraphics[width=0.5\textwidth]{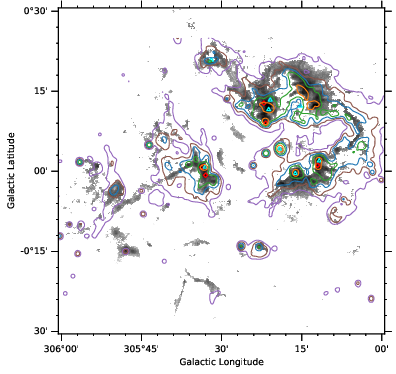}
            \caption{\ce{^{13}CO} $J=3\text{-}2$ moment 0 map with GLIMPSE 8$\,\mu\si{\metre}$ contours corresponding to 84(purple), 123(brown), 178(blue), 260(green) and 550.(orange)$\, \si{\mega\jansky\per\steradian}$ overlaid on top. The HII and UC HII regions are marked by cyan triangles and red diamonds respectively.}
            \label{fig:13CO_8um_feedback_contour}
            \end{figure}

 To investigate possible causes for the trends we find in the statistical moments of the stacked spectra, we plotted the contours of 8$\,\mu\si{\metre}$ flux on top of \ce{^{13}CO} integrated intensity map. The relevant 8$\,\mu\si{\metre}$ flux values at which the statistical moments change directions are 84, 123, 178, 260 and 378$\,\si{\mega\jansky\per\steradian}$. Fig. \ref{fig:13CO_8um_feedback_contour} shows these contours overlaid on top of the \ce{^{13}CO} moment 0 map. Almost all of the dense gas traced by the \ce{^{13}CO} is contained inside the $123\,\si{\mega\jansky\per\steradian}$ contour. We expect the feedback from the central clusters to be deposited on the surface of the dense gas. As we examine the higher flux contours, they appear to surround the HII and UC HII regions in the complex \citep{hindson12}. It is possible that the high column density gas surface interacting with the feedback from the central clusters is responsible for the observed increase in skewness and Kurtosis and the decrease in the standard deviation at $\sim 100 \, \si{\mega\jansky\per\steradian}$, as this is where we expect the feedback from the central stars to be deposited. As the 8$\,\mu\si{\metre}$ flux increases, the gas is impacted mostly by the stellar winds and ionizing radiation from embedded HII and UC HII regions which appears to be responsible for the increase in skewness as well as kurtosis and quite interestingly, the decrease in standard deviation of the stacked line profiles. The \ce{^{13}CO} emission from our observation suffers from a lack of completeness as we do not sample a lot of low column density gas for a given 8$\,\mu\si{\metre}$ flux. Most of the spectra outside the masks are dominated by noise. In addition, the wings of the line profiles are also dominated by noise in many cases even when they are significantly detected in \ce{^{12}CO}. Since, these wings correspond to the gas being expelled, it is difficult to trace them using \ce{^{13}CO}. Due to these reasons, it was not possible to repeat the stacked spectra analyses with \ce{^{13}CO} lines to gain any further insights into the causes of the observed trends. Examining the literature, we would like to emphasize that apparently this kind of analysis has not yet been performed, neither on any actual nor on simulated observations of molecular cloud complexes undergoing feedback. So far, this is a stand alone result and it needs to be seen if these observed trends in stacked spectra (if real) are unique to G305 or occur in other GMCs as well before making any meaningful speculations on the causes of these trends.

\section{Summary \label{sec:conc}}
    We observed the G305 star forming giant molecular cloud with the APEX telescope in the \ce{^{12}CO} and \ce{^{13}CO} $J=3\text{-}2$ transitions in order to study the effects of feedback from the hot, luminous stars at the center of the complex on the molecular gas. We summarize our finding below:
    
    \begin{itemize}
     \item  The central region of the complex has been cleared out of most of the high column density gas as traced by the \ce{^{13}CO} emission. The calculations of energy input from the visible stars in the complex showed that they have enough energy input to drive the size of the cavity observed if the electron density ($n_e$) in the region is less than $500\,\si{\per\cubic\cm}$. A sequential formation of Danks 2 followed by Danks 1 and the other stars in the complex can explain the size of the observed cavity.
     \item \ce{^{12}CO} excitation temperature and \ce{^{13}CO} column density maps of the region were produced under the assumption of LTE. Ratio maps of the rotational excitation were also made using \ce{^{13}CO} $J=2\text{-}1$ data from SEDIGISM and our \ce{^{13}CO} $J=3\text{-}2$ data. The validity of LTE assumption was also tested and it was concluded that regions with simultaneous emission from both \ce{^{12}CO} and \ce{^{13}CO} are most likely in LTE.
     \item  Excitation temperature maps as well as ratio maps of the rotational excitation show that the feedback is being deposited in a narrow region at the edge of the dense gas facing the central stellar complex heating it up. The gas then shows a decline in temperature as one moves away from the center. The column density also shows a marked increase at the edge of the denser gas, but unlike the excitation temperature, does not always decrease drastically as one moves away from the center.
     \item Line profiles along two directions were chosen at random to test if the energy input from the stars is responsible for this increased excitation of the gas. For the profile towards the north of the complex $\sim 61\%$ of the radiation pressure from the nearby stars is being used up to heat and compress the gas at the surface of the cloud; whereas for that towards the south of the complex, $<14.5\%$ of the input radiation pressure from the star is going towards heating the gas.
     \item The \textit{GLIMPSE} 8$\,\mu\si{\metre}$ flux, which is dominated by FUV-excited PAH emission, was used as a proxy to the feedback strength. The regions with higher 8$\,\mu\si{\metre}$ flux have higher median excitation temperatures, \ce{^{13}CO} column density as well as a higher median \ce{^{13}CO} $J=3\text{-}2/2\text{-}1$ ratio.
     \item Investigating the impact of feedback on the dynamics of the gas showed that the centroid velocity probability distribution function of the pixels in the region showed exponential wings indicative of turbulence driven by strong stellar feedback.
     \item This was followed by stacking the spectra of the pixels and plotting the average stacked profiles. On assuming a certain 8$\,\mu\si{\metre}$ flux threshold, the stacked spectra with 8$\,\mu\si{\metre}$ flux above this threshold (assumed to indicate stronger feedback) showed systematically more skewed line profiles than the stacked spectra representing regions with 8$\,\mu\si{\metre}$ flux less than the threshold. The stacked spectra of regions with stronger feedback on an average had narrower but much more winged profiles when compared to those from regions with weaker feedback. This positive skew is most likely indicative of parts of the cloud complex being pushed away from us, which results in a positive wing of the overall stacked spectra. We also note that the standard deviation and the kurtosis of the stacked profiles show opposing trends when plotted as a function of 8$\,\mu\si{\metre}$ flux.
    \end{itemize}

\begin{acknowledgements}
We thank the staff of the APEX telescope for their assistance in observations. We also thank the anonymous referee whose invaluable suggestions have significantly improved the quality of the paper. This work acknowledges support by The Collaborative Research Council 956, sub-project A6, funded by the Deutsche Forschungsgemeinschaft (DFG). DC acknowledges support by the German \emph{Deut\-sche For\-schungs\-ge\-mein\-schaft, DFG\/} project number SFB956A. Parts of this work are based on observations made with the Spitzer Space Telescope, which is operated by the Jet Propulsion Laboratory (JPL), California Institute of Technology under a contract with NASA. This publication also made use of data products from the Midcourse Space Experiment. Processing of the data was funded by the Ballistic Missile Defense Organization with additional support from NASA Office of Space Science. This research has also made use of the NASA/ IPAC Infrared Science Archive, which is operated by the JPL, under contract with NASA.
\end{acknowledgements}

\bibliographystyle{aa} 
\bibliography{allpapers} 

\appendix
\onecolumn

\section{Noise and Offset Details \label{app:a}}
    \begin{table}[h]
        \renewcommand*{\arraystretch}{1.4}
        \centering
        \caption{Position and Noise towards off-positions used in otf-mapping of G305.}
        \begin{tabular}{|c|c|c|c|}
            \hline
            \textbf{RA [J2000]} & \textbf{Dec [J2000]} & \textbf{Freq.-Range [GHz]} & \textbf{Noise rms [K]}\tablefootmark{a}\\
            \hline
            13:08:23.90 & -60:48:13.0 & 330.50-330.65 & 1.001e-2\\
            &  & 345.70-345.85 & 1.068e-2\\
            13:10:52.20 & -64:47:37.0 & 330.50-330.65 & 1.984e-2\\
            &  & 345.70-345.85 & 1.884e-2\\
            13:20:12.00 & -64:42:01.0 & 330.50-330.65 & 2.116e-2\\
            &  & 345.70-345.85 & 2.077e-2\\
            13:16:33.30 & -60:43:19.0 & 330.50-330.65 & 2.357e-2\\
            &  & 345.70-345.85 & 2.176e-2\\
            \hline
        \end{tabular}
        \tablefoot{\tablefoottext{a}{The rms noise was obtained by averaging over all 7 pixels, albeit being at different offsets with respect to the position mentioned in the table.}}
    \end{table}

    \begin{figure}[h]
        \centering
        \parbox {0.45\textwidth}{
            \includegraphics[width=0.4\textwidth]{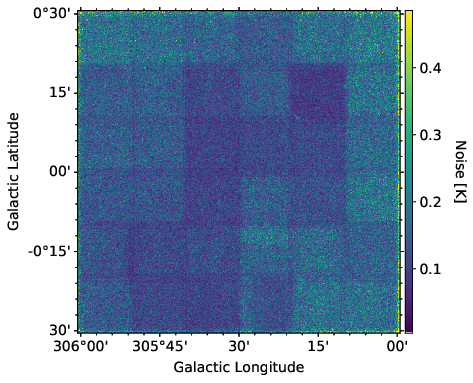}}
        \\
        \begin{minipage}{0.45\textwidth}
            \includegraphics[width=0.9\textwidth]{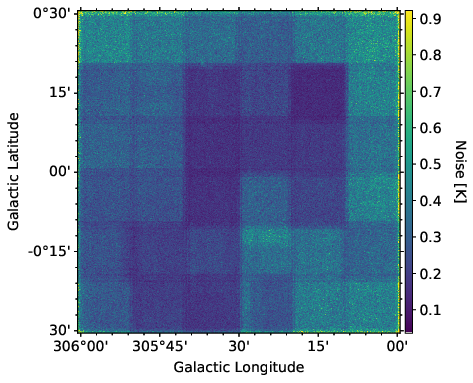}
        \end{minipage}
        \caption{Pixel wise noise maps of \ce{^{12}}CO(3-2) (\textit{top}) and \ce{^{13}CO}(3-2) (\textit{bottom}) of the G305 GMC complex.}
        \label{fig:noise_map}
    \end{figure}

\end{document}